%% file: main.tex
\documentclass[prd,showpacs,superscriptaddress,twocolumn,floatfix,preprintnumbers,altaffilletter]{revtex4-1}
\usepackage{booktabs}
\usepackage{amssymb, amsmath}
\usepackage[utf8]{inputenc}
\usepackage{epsfig}
\usepackage{xcolor}
\usepackage{graphicx}
\usepackage[colorlinks=true]{hyperref}
\usepackage[caption=false]{subfig}
\usepackage{xspace}

\newcommand{\ud}{\mathrm{d}}
\newcommand{\n}{\frac{\ud^2 \cal N}{\ud \ell \ud V}}
\newcommand{\orad}{\mathrm{\Omega_{rad}}}

\newcommand{\chir}{\chi_\mathrm{{rad}}}
\newcommand{\chim}{\chi_\mathrm{{mat}}}
\newcommand{\uH}{\mathrm{H_0}}

\begin{document}

\title{Constraints on cosmic strings using data from the third Advanced LIGO--Virgo observing run}

\let\mymaketitle\maketitle
\let\myauthor\author
\let\myaffiliation\affiliation
\author{The LIGO Scientific Collaboration, the Virgo Collaboration and the KAGRA Collaboration}

\date[\relax]{compiled \today}

\begin{abstract}
  We search for gravitational-wave signals produced by cosmic strings in the Advanced LIGO and Virgo full O3 data set. Search results are presented for gravitational waves produced by cosmic string loop features such as cusps, kinks and, for the first time, kink-kink collisions.
  A template-based search for short-duration transient signals does not yield a detection. 
  We also use the stochastic gravitational-wave background energy density upper limits derived from the O3 data to constrain the cosmic string tension, $G\mu$, as a function of the number of kinks, or the number of cusps, for two cosmic string loop distribution models.
  Additionally, we develop and test a third model which interpolates between these two models.
  Our results improve upon the previous LIGO-Virgo constraints on $G\mu$ by one to two orders of magnitude depending on the model which is tested. In particular, for one loop distribution model, we set the most competitive constraints to date, $G\mu\lesssim 4\times 10^{-15}$.
\end{abstract}



\maketitle

\input{Introduction}


\input{GW_loops}

\input{Burst_search}

\input{Stoch_search}

\input{Constraints}

\input{Conclusions}

{\em Acknowledgments.}--
\input{P2000488_v3.tex}

This document has been assigned the LIGO document number LIGO-P2000506.

\appendix
\input{Appendix_model}

\input{Appendix_burst}

\bibliography{references.bib}

\let\author\myauthor
\let\affiliation\myaffiliation
\let\maketitle\mymaketitle
\title{Authors}
\pacs{}

\input{LSC-Virgo-KAGRA-Authors-Aug-2020-prd}
\newpage
\maketitle

\end{document}

%% file: Introduction.tex
\section{Introduction}
\label{sec:intro}

The Advanced LIGO~\cite{0264-9381-32-7-074001} and Advanced Virgo~\cite{0264-9381-32-2-024001} detectors have opened a new channel to observe the Universe through the detection of gravitational waves.
In their first three observing runs (O1, O2, and the first half of O3) the LIGO Scientific Collaboration and the Virgo Collaboration have reported the detection of 50 candidate gravitational-wave events from compact binary coalescences ~\cite{Abbott:2020niy}.
These detections have yielded important information on the population properties of these compact binary sources~\cite{Abbott:2020gyp}.
In the future, ground-based detectors may discover new sources of gravitational waves~\cite{Riles20131}, some of which could probe the physics of the early Universe. Cosmic strings~\cite{Kibble:1976sj} belong to this category of sources.
The third observing run (O3) started on April 1, 2019, and ended on March 27, 2020, and we use the data from the LIGO-Hanford (H1), LIGO-Livingston (L1) and Virgo (V1) interferometers to place constraints on cosmic strings. These constraints are reported in this letter.

Cosmic strings are line-like topological defects - analogues of vortices in different condensed matter systems - which are formed from spontaneous symmetry breaking phase transitions (with the additional condition that the vacuum manifold has non-contractible closed curves~\cite{Kibble:1976sj,Vilenkin:2000jqa,Hindmarsh:1994re,Vachaspati:2015cma}). In cosmology, such phase transitions may have occurred at grand unification~\cite{Jeannerot:2003qv}, corresponding to an energy scale of about $ 10^{16} ~\mathrm{GeV}$, and more generally at lower energy scales.
Thus cosmic strings, through their different observational predictions, offer a tool to probe particle physics beyond the Standard Model at energy scales much above the ones reached by accelerators.
In particular, the production of gravitational waves by cosmic strings~\cite{Vachaspati:1984gt,Sakellariadou:1990ne} is one of the most promising observational signatures that can be accessed by ground-based detectors.

At these energy scales, the width of a string is negligible relative to its characteristic size, so that the cosmic string is well described by the Nambu-Goto action.
Nambu-Goto strings~\cite{Vilenkin:2000jqa} are parameterized by a dimensionless quantity: the string tension $G\mu$ related to the energy scale $\eta$ at which cosmic strings are formed, $G\mu\sim(\eta/M_{\rm Pl})^2$, where $G$ stands for Newton's constant, $M_{\rm Pl}$ is the Planck mass, $\mu$ denotes the string linear mass density~\footnote{Cosmic superstrings~\cite{Polchinski:2004yav}, the analogues of cosmic strings arising from string theory, are characterized also by the intercommutation probability which can take values between $10^{-3}$ and 1 for fundamental superstrings (F-strings) and between $10^{-1}$ and 1 for D-branes extended in one macroscopic dimension (D-strings). In our present study we concentrate on field theoretical objects~\cite{Sakellariadou:2009ev}, and in particular Nambu-Goto strings with intercommutation probability of order one.}, and we set the speed of light $c=1$.
Within the context of cosmology, a cosmic string network relaxes towards a scaling solution -- a self-similar, attractor solution in which typical loop lengths are proportional to cosmic time, namely they scale with the Hubble radius. Super-horizon (infinite) strings reach a scaling solution~\cite{BB1988,Allen:1990tv,Sakellariadou:1990nd} by losing energy through the formation of sub-horizon (loop) strings, which consequently lead to a cascade of smaller loops eventually decaying through emission of gravitational waves~\cite{Vilenkin:1981bx,Hogan:1984is,Sakellariadou:1990ne}.
In this paper we focus on the gravitational waves emitted by the network of loops.
The length distribution of loops will therefore be crucial in determining the gravitational-wave signatures from cosmic strings.
We consider different models for the loop distribution, each of which has been studied in the literature, and whose differences arise from different modeling of the production and cascade of loops from the infinite string network.

Cosmic string loops oscillate periodically in time, emitting gravitational waves with power~\cite{Vachaspati:1984gt} $P_{\rm gw}=\Gamma_{\rm d} G\mu^2$ and decay in a lifetime $\ell/\gamma_{\rm d}$, where $\Gamma_{\rm d}$ is a numerical factor, $\ell$ is the invariant  loop length and $\gamma_{\rm d}=\Gamma_d G\mu$ is the gravitational-wave length scale measured in units of time~\footnote{Super-horizon cosmic strings also emit gravitational waves, due to their small-scale structure resulting from string intercommutations~\cite{Sakellariadou:1990ne,Hindmarsh:1990xi,Matsui:2016xnp}.}.
The high-frequency ($f\ell\gg 1$, where $f$ denotes frequency) gravitational-wave spectrum of an oscillating loop is dominated by bursts emitted by string features called cusps and kinks~\cite{Damour:2000wa,Damour:2001bk,Damour:2004kw}. Cusps~\footnote{We also include the so-called pseudocusps~\cite{Stott:2016loe}, defined as cuspy features moving with a velocity close to the speed of light.} are points on the string that briefly travel at the speed of light; they are generic features for smooth loops. Kinks are discontinuities in the tangent vector of the string that propagate at the speed of light. They appear in pairs as the result of collisions between two cosmic strings and are chopped off when a loop forms, hence a loop can contain any integer number of kinks. Numerical simulations of Nambu-Goto strings have shown that kinks accumulate over the cosmological evolution~\cite{BB1988,Allen:1990tv,Sakellariadou:1990nd}, while the number of cusps per loop is yet undetermined.

Cusps are short-lived and produce beamed gravitational waves in the forward direction of the cusp, while kinks propagate around the string, creating gravitational waves with a fan-like emission (like a lighthouse). Additionally, the collision of two kinks is expected to radiate gravitational waves isotropically. In this paper, we report on searches for gravitational waves produced by cusps, kinks and kink-kink collisions, using O3 LIGO--Virgo data. In addition to distinct individual bursts, the  incoherent superposition of weaker gravitationa-wave bursts  from  cosmic  strings produced over the history of the Universe would create a stochastic gravitational-wave background~\cite{Siemens:2006yp,Damour:2004kw}.

The gravitational-wave emission from cosmic string loops is introduced in Sec.~\ref{sec:GW_loops}. We consider two simulation-based models~\cite{Blanco-Pillado:2013qja,Lorenz:2010sm} (labeled \textbf{A} and \textbf{B}) for the distribution of cosmic string loops. In addition, we develop a third model (labeled \textbf{C}) which interpolates between the other two models. From these, we derive gravitational-wave burst rates and the dimensionless energy density spectra
in Sec.~\ref{sec:GW_loops}. Individual gravitational-wave bursts are searched in O3 data with a dedicated analysis presented in Sec.~\ref{sec:burst_search}. The incoherent superposition of bursts from cusps, kinks and kink-kink collisions, produces a stationary and nearly Gaussian stochastic background of gravitational waves. We search O3 data for this background and the results, detailed in~\cite{O3_isotropic}, are summarized in Sec.~\ref{sec:stoch_search}. Both the burst and stochastic background searches yield no detections. Combining their sensitivities, we constrain two cosmic string parameters in Sec.~\ref{sec:constraints}: the string tension $G\mu$ and the number of kinks per loop.

%% file: GW_loops.tex
\section{Gravitational waves from cosmic string loops}
\label{sec:GW_loops}

Gravitational waves are produced by cusps, kinks and kink-kink collisions on cosmic string loops.
The waveforms of the strain have been calculated in~\cite{Damour:2001bk,Damour:2004kw,Damour:2000wa}.
For a loop of length $\ell$ at redshift $z$, they are power-law functions in the frequency domain for the strain~\cite{Abbott:2017mem}:
\begin{eqnarray}
  h_i(\ell,z,f) = A_i(\ell, z) f^{-q_i} ,
  \label{eq:h}
\end{eqnarray}
where $i=\{{\rm c},{\rm k},{\rm kk}\}$ identifies the cusp, kink, and kink-kink collision cases. The power-law indices are $q_{\rm c}=4/3$, $q_{\rm k}=5/3$, and $q_{\rm kk}=2$ and the signal amplitude $A_i$ is given by~\cite{Damour:2001bk}
\begin{eqnarray}
  A_i(\ell, z) =  g_{1, i} \frac{G \mu\ \ell^{2-q_i}}{(1+z)^{q_i-1} r(z)},
  \label{eq:Aq}
\end{eqnarray}
where $r(z)$ is the comoving distance to the loop. In the following we adopt the cosmological model used in~\cite{Abbott:2017mem}; it is encoded in three functions $\varphi_r(z)$, $\varphi_V(z)$, and $\varphi_t(z)$ (see Appendix A of \cite{Abbott:2017mem}). The proper distance, the proper volume element and the proper time are given by $r(z)=\varphi_r(z)/H_0$, $\mathrm{d}V(z)=\varphi_V(z)/H_0^3 \mathrm{d}z$, and $t(z)=\varphi_t(z)/H_0$ respectively, where $H_0=67.9\ {\rm km\ s^{-1}\ Mpc^{-1}}$~\cite{Planck_2015} is the present value of the Hubble constant. The prefactor $g_{1,i}$ takes three different numerical values~\cite{Jenkins:2018lvb}: $g_{1,{\rm c}}=8/\Gamma^2(1/3)\times (2/3)^{2/3}\approx 0.85$, $g_{1,{\rm k}}=2\sqrt{2}/\pi/\Gamma(1/3)\times (2/3)^{2/3}\approx 0.29$, and $g_{1,{\rm kk}}=1/\pi^2\approx 0.10$, where $\Gamma$ is the Gamma function~\cite{gradshteyn2007}.

Cusps and kinks emit gravitational waves in highly concentrated beams. Cusps are transient and produce a beam along a single direction, while kinks propagate around the loop, beaming over a fan-like range of directions. The beam opening angle is
\begin{eqnarray}
  \theta_m = (g_2 f(1+z) \ell)^{-1/3},
  \label{eq:thetam}
\end{eqnarray}
where $g_2= \sqrt{3}/4$~\cite{Jenkins:2018lvb} is a numerical factor. To guarantee self-consistency, we require that $\theta_m<1$, which is equivalent to setting a lower limit on the frequency for a fixed loop length. For kink-kink collisions the gravitational-wave emission is isotropic~\cite{Binetruy:2009vt}.

The burst rate of type $i$, per unit loop size and per unit volume, can be decomposed into four factors:
\begin{equation}
    \frac{\ud R_i}{\ud \ell \ud V} =\frac{2}{\ell}N_i \times n(\ell,t) \times \Delta_i \times (1+z)^{-1}.
\end{equation}
The first factor accounts for an average of $N_i$ gravitational-wave burst events of type $i$ produced per loop oscillation time periodicity $\ell/2$. The second factor stands for the number of loops per unit loop size and per unit volume
\begin{equation}
n(\ell,t) = \n(\ell,t).
\end{equation}
The third factor, $\Delta_i$, reflects that only a fraction of burst events can be effectively detected due to the beamed emission of gravitational waves with respect to the $4\pi$ solid angle. The gravitational-wave emission within a cone for cusps, a fan-like range of directions for kinks and all directions for kink-kink collisions can be conveniently absorbed into a single beaming fraction expression: $\Delta_i=(\theta_m/2)^{3(2-q_i)}$. Finally the last factor shows that the burst emission rate is red-shifted by $(1+z)^{-1}$.

The burst rate at redshift $z$ is then obtained by integrating over all loop sizes,
\begin{equation}
\frac{\ud R_i}{\ud z}=\frac{\varphi_V(z)}{H_0^3(1+z)}\int_{\ell_{\text{min}}}^{\ell_{\text{max}}}\ud\ell \frac{2 N_i}{\ell} n(\ell,t) \Delta_i.
  \label{eq:drdz_ell}
\end{equation}
Introducing the dimensionless loop size parameter $\gamma \equiv \ell/t$, Eq.~\ref{eq:drdz_ell} reads:
\begin{eqnarray}
  \frac{\ud R_i}{\ud z}(z,f) &=& \frac{\phi_V(z)}{H_0^3(1+z)}\nonumber \\
  &&\times \int_{\gamma_{\text{min}}(z,f)}^{\gamma_{\text{max}}(z)} \ud \gamma \frac{2 N_i}{\gamma} n(\gamma, z) \Delta_i(\gamma, z, f).
  \label{eq:drdz_gamma}
\end{eqnarray}
The upper bound of the integral, $\gamma_{\text{max}}(z)$, is derived by requiring the loop size to be smaller than the horizon size, i.e., $\gamma_{\text{max}} = 2$ and 3 for radiation and matter dominated universes, respectively~\cite{Abbott:2017mem}. The lower bound, $\gamma_{\text{min}}$, corresponds to the fundamental frequency of a loop, i.e., $2/\ell$, leading to $\gamma_{\text{min}}(z, f) = 2/[f(1+z)\varphi_t(z)/H_0]$.

We consider two analytical models, labeled {\bf A}~\cite{Blanco-Pillado:2013qja} and {\bf B}~\cite{Lorenz:2010sm}, to describe the distribution of cosmic string loops, $n(\gamma,z)$, in a scaling regime, within a Friedmann-Lema\^{i}tre-Robertson-Walker metric. These models were respectively dubbed M=2 and M=3 in~\cite{Abbott:2017mem}. In model {\bf A} the
 number of long-lived non-self-intersecting loops of invariant length $\ell$ per unit volume per unit time formed at cosmic time $t$, is directly inferred from Nambu-Goto simulations of cosmic string networks in the radiation and matter eras. Model {\bf B} is based on a different Nambu-Goto string simulation~\cite{Ringeval:2005kr}. In this model
the distribution of non-self-intersecting scaling loops is the extracted quantity. Within model {\bf B}, loops are formed at all sizes following a power-law specified by a parameter taking different values in the radiation and matter eras, while the
scaling loop distribution is cut-off on small scales by the gravitational back-reaction scale. There is a qualitative difference between these two models, since in the latter, tiny loops are produced in a much larger amount than in the former. In addition, we will use a
new model, based on~\cite{Auclair:2019zoz} and labeled {\bf C}, which extends and encompasses both models {\bf A} and {\bf B}.
Similarly to model \textbf{B}, model \textbf{C} assumes that the
scaling loop distribution is a power-law, but leaves its slope unspecified.
Given the wide parameter space opened by
model \textbf{C}, we will select two samples: models \textbf{C-1} and \textbf{C-2}.
Model \textbf{C-1} (respectively \textbf{C-2}) reproduces qualitatively the loop production function of model \textbf{A} (resp. \textbf{B}) in the radiation era and the loop production of model \textbf{B} (resp. \textbf{A}) in the matter era.
We expect the addition of these two models to showcase intermediate situations in between the two simulation-inferred models \textbf{A} and \textbf{B}.
The loop distribution functions $n(\gamma,z)$ for the three models are given in the Supplemental Material~\footnote{See Supplemental   Material for more descriptions on the cosmic string loop distributions, and the burst analysis pipeline}.

For models {\bf A}, {\bf B} and {\bf C}, the contributions from cusps, kinks and kink-kink collisions to the gravitational-wave emission must be considered altogether. Indeed, the dimensionless decay constant $\Gamma_{\rm d}$ of a cosmic string, driving the loop size evolution, can be decomposed into three contributions:
\begin{eqnarray}
  \label{Gamma}
  \Gamma_{\rm d} &\equiv& {P_{\rm gw}\over G\mu^2}=\sum_i {P_{{\rm gw}, i}\over G\mu^2}\nonumber\\
  &=& N_{\rm c}\frac{3\pi^2g_{1, {\rm c}}^2}{(2 \delta)^{1/3}g_2^{2/3}}+N_{\rm k}\frac{3\pi^2g_{1, {\rm k}}^2}{(2 \delta)^{2/3}g_2^{1/3}}
  +N_{\rm kk}2\pi^2g_{1, {\rm kk}}^2,\nonumber\\
\end{eqnarray}
where $\delta = \text{max}(1, 1/(2 g_2))$, since the gravitational-wave frequency cannot be smaller than the fundamental frequency of the loop, $2/\ell$, while the condition $\theta_m<1$ for cusps and kinks imposes $ f > 1/(\ell g_2)$.
Parameters $N_{\rm c}, N_{\rm k}$ are respectively the average number of cusps and kinks per oscillation. The number of kink-kink collisions per oscillation $N_{\rm kk}$ is $N_{\rm kk} \approx N_{\rm k}^2/4$ in the limit of large $N_{\rm k}$. While this equation is only an approximation when $N_{\rm k}$ is order unity, the kink-kink contribution is very small in this case and the error would hardly affect our results. On the other hand, it is clear that the kink-kink collision quickly dominates the gravitational-wave production when the number of kinks increases. In this analysis we fix $N_{\rm c}$ to be $1$; we comment later on the effects of increasing the number of cusps. The only free parameter is the number of kinks $N_{\rm k}$; we consider  $N_{\rm k} = 1,\dots, 200$;, with the upper limit being motivated by numerical simulations of string loops that favor $\Gamma_{\rm d}\sim 50 $~\cite{Allen:1991bk}.

\begin{figure*}
\includegraphics[width=0.48\textwidth]{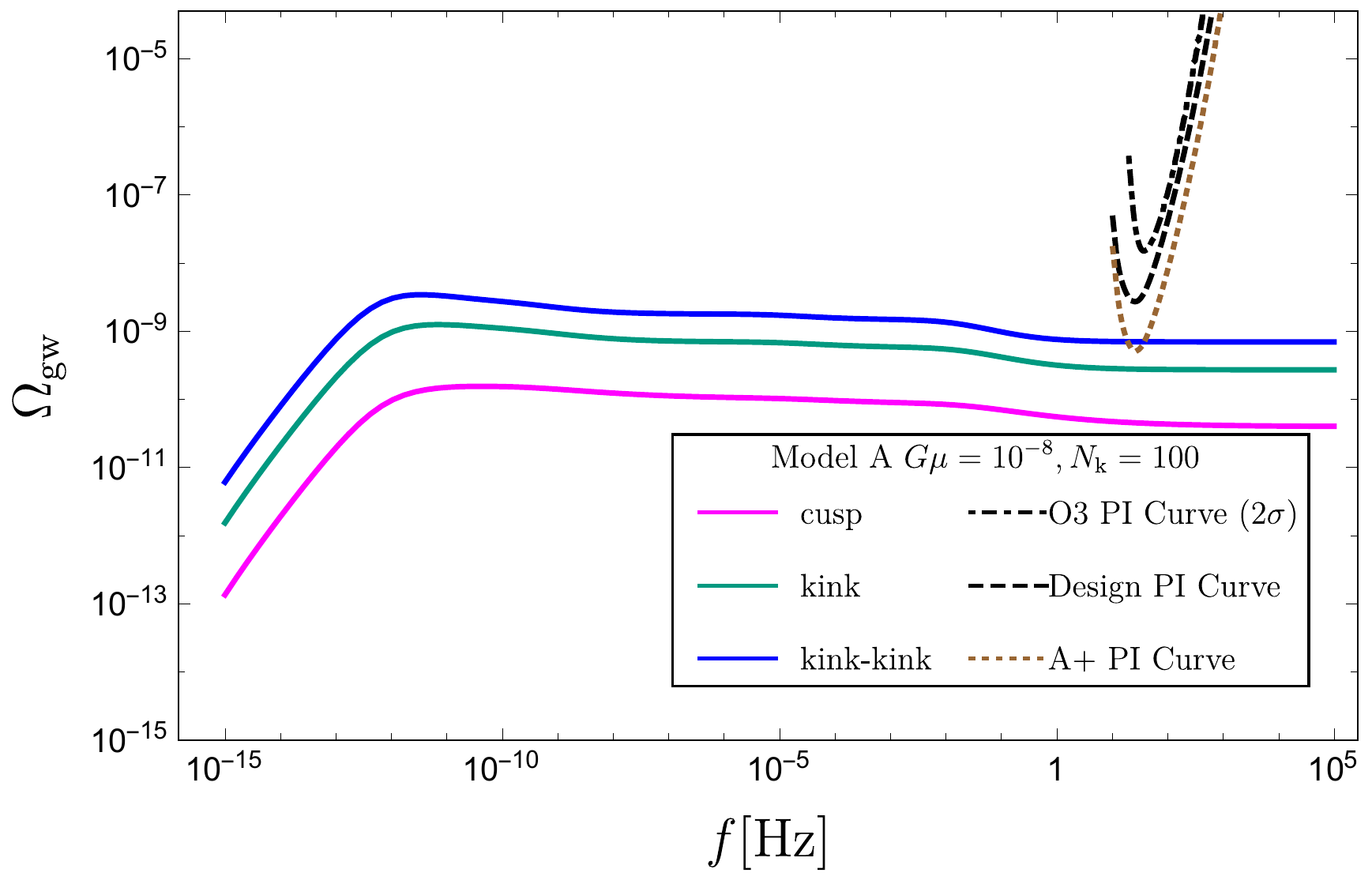}
\includegraphics[width=0.48\textwidth]{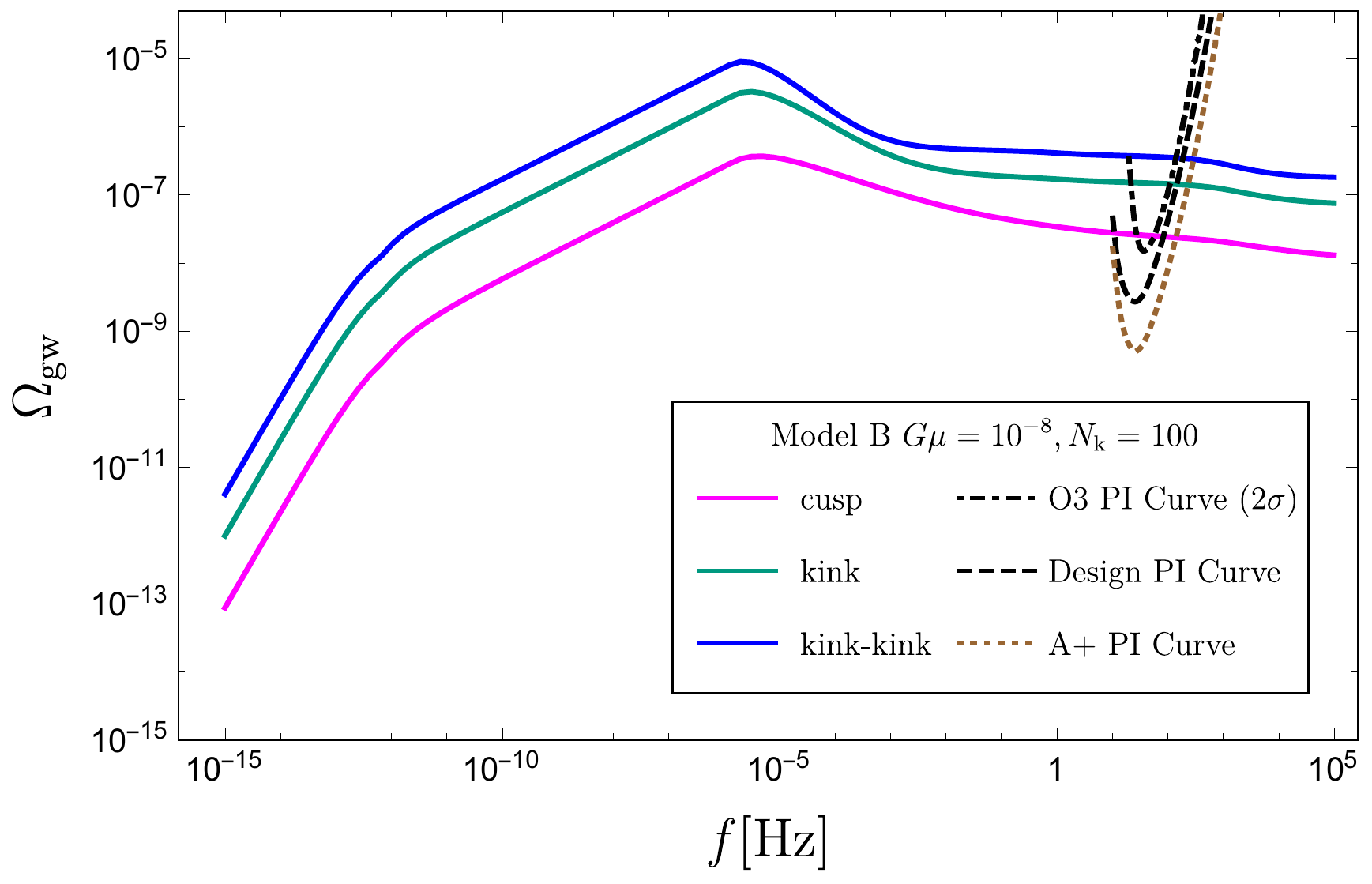}
\\
\includegraphics[width=0.48\textwidth]{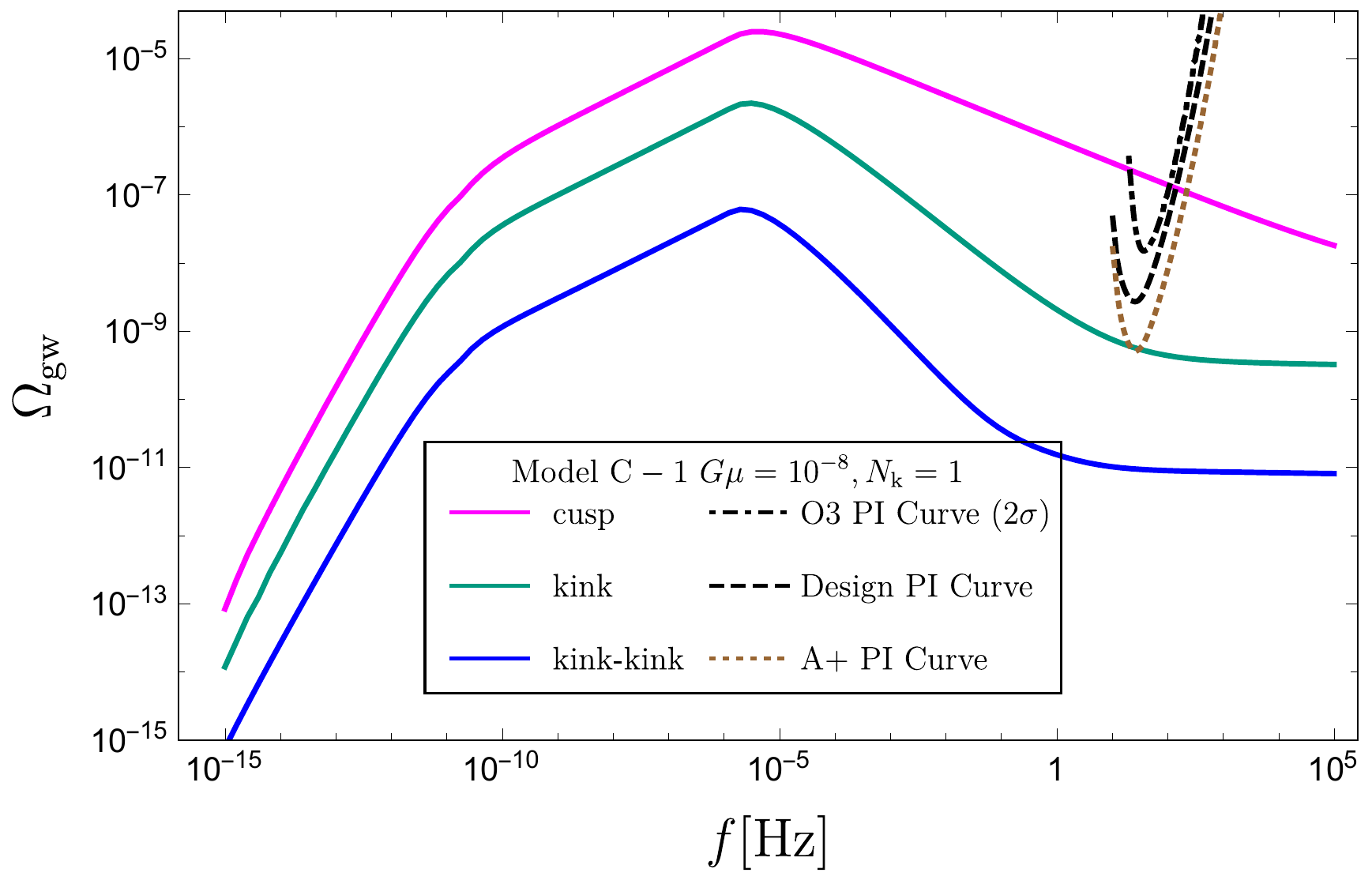}
\includegraphics[width=0.48\textwidth]{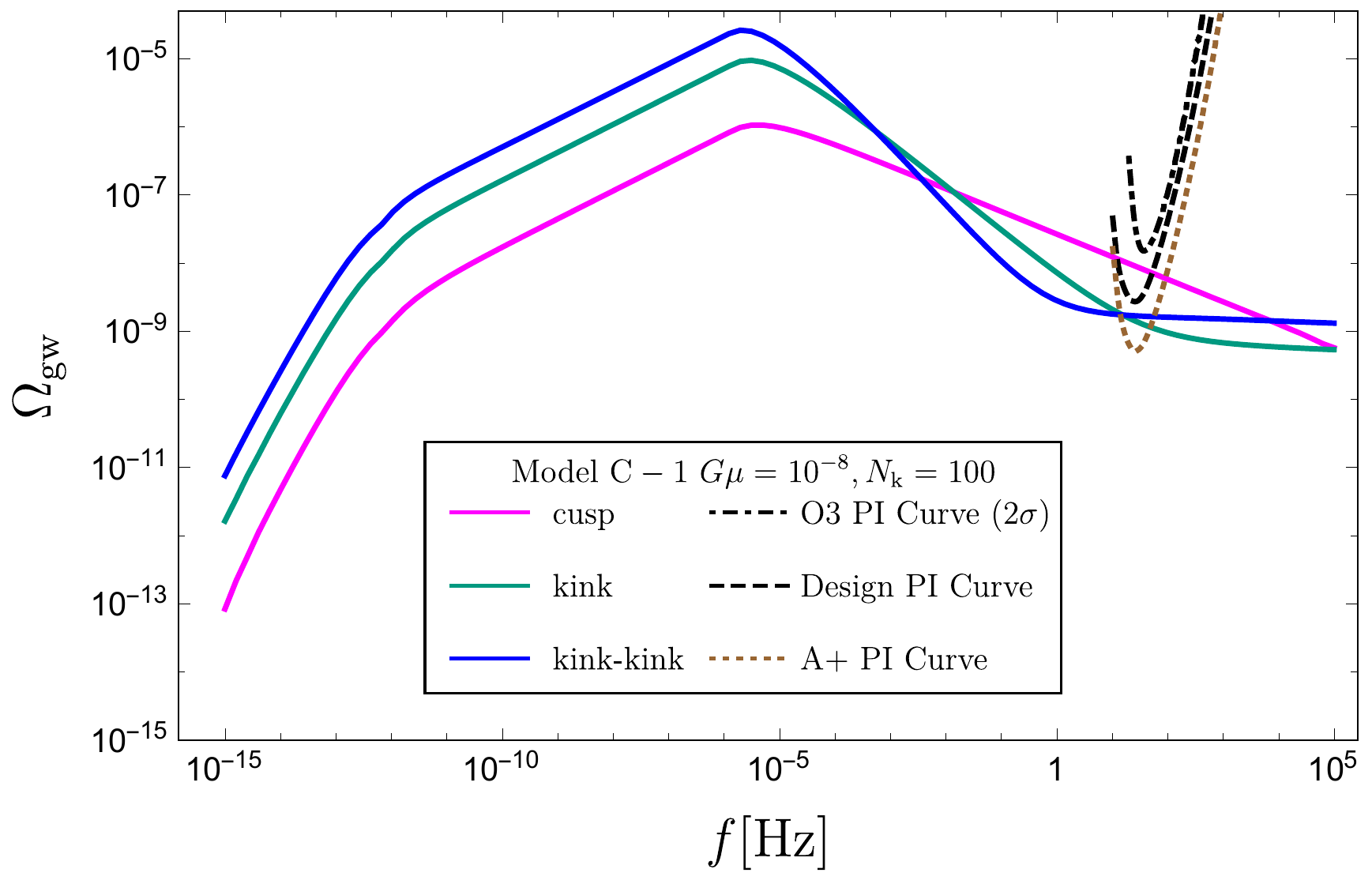}
\caption{Predictions of the gravitational-wave energy density spectra using different models for the loop distribution function $n(\gamma,z)$ and varying the number of kinks per loop oscillation $N_{\rm k}$. The string tension $G\mu$ is fixed to $10^{-8}$.
  Top-left: model {\bf A}, $N_{\rm k}=100$.
  Top-right: model {\bf B}, $N_{\rm k}=100$.
  Bottom-left: model {\bf C-1}, $N_{\rm k}=1$.
  Bottom-right: model {\bf C-1}, $N_{\rm k}=100$.
  For model {\bf C-1}, we use the following model parameters (see Supplemental Material): $\chi_{\rm rad}=0.45$, $\chi_{\rm mat}=0.295$, $c_{\rm rad}=0.15$, $c_{\rm mat}=0.019$; the subscripts refer to the radiation and matter eras, respectively. We also show the energy density spectra of the three different components and 2-$\sigma$ power-law integrated (PI) curves~\cite{PhysRevD.88.124032} for the O3 isotropic stochastic search~\cite{O3_isotropic}, and projections for the HLV network at design sensitivity, and the A+ detectors~\cite{Aasi:2013wya}.
}\label{spec}
\end{figure*}

The incoherent superposition of bursts from loops with all possible sizes through the history of the Universe produces a stochastic gravitational wave background~\cite{Christensen:2018iqi}. The normalized energy density of which is defined as
\begin{equation}
  \Omega_{\rm GW}(f) = \frac{f}{\rho_c} \frac{{\rm d} \rho_{\rm GW}}{{\rm d} f} ~,
  \label{sb}
\end{equation}
where $\rho_{\rm c}=3 H_0^2 c^2/(8\pi G)$. The spectrum of the stochastic gravitational wave background is~\cite{Olmez:2010bi}
\begin{eqnarray}
  \Omega_{\text{GW}}(f) = \frac{4\pi^2}{3 H_0^2} f^3 \sum_i \int \ud z \int \ud \ell ~ h_i^2 \times \frac{\ud^2 R_i}{\ud z \ud \ell} .
  \label{eq:Omega}
\end{eqnarray}
In Fig.~\ref{spec} we show examples of gravitational-wave spectra calculated with Eq.~\ref{eq:Omega}. The two plots at the top are derived from model {\bf A} and {\bf B} with $N_k \gg 1$. The dominant contribution comes from kink-kink collisions. The lower plots show gravitational-wave spectra taking $N_k=1$ (left) and $N_k=100$ (right) and are derived from model {\bf C} with a given set of parameters (see Supplemental Material), i.e., $\chi_{\rm rad}=0.45$, $\chi_{\rm mat}=0.295$, $c_{\rm rad}=0.15$, and $c_{\rm mat}=0.019$; the subscripts refer to matter and radiation eras, respectively. When $N_{\rm k}$ is large, the dominant contribution to the spectrum depends on the frequency band, which is a unique feature in this model. In this study, we ignore the suppression of the gravitational waves from cusps due to the primordial black hole production as pointed out in~\cite{Jenkins:2020ctp}. Including such an effect leads to lower spectrum amplitudes when $N_{\rm k}$ is small and consequently reduces the sensitivity to cosmic string signals.
In Fig.~\ref{spec} we also show the $2 \sigma$ power-law integrated (PI) curves~\cite{PhysRevD.88.124032} indicating the integrated
sensitivity of the O3 search~\cite{O3_isotropic}, along with projections for 2 years of the Advanced LIGO--Virgo network at design sensitivity, and the envisioned upgrade of Advanced LIGO, A+~\cite{Aasi:2013wya}, sensitivity after
2 years, assuming a 50\% duty cycle.

%% file: Burst_search.tex
\section{Burst Search}\label{sec:burst_search}

The O3 data set is analyzed with a dedicated burst search algorithm previously used to produce LIGO--Virgo results~\cite{Abbott:2019prv,Abbott:2017mem,Aasi:2013vna}. The burst analysis pipeline, as well as its O3 configuration, is described in the Supplemental Material. The search can be summarized into three analysis steps.
First, we carry out a matched-filter search using the cosmic string waveform in Eq.~\ref{eq:h}.
Then, resulting candidates are filtered to retain only those detected in more than one detector within a time window accounting for the difference in the gravitational-wave arrival time between detectors. Finally, double- and triple-coincident events are ranked using a likelihood function $\Lambda(x)$, where $x$ is a set of parameters used to discriminate true cosmic string signals from noise~\cite{Cannon:2008zz}. The burst search is performed separately for cusps, kinks and kink-kink collision waveforms, integrating $T_{\mathrm{obs}}=273.5$\,days of data when at least two detectors are operating simultaneously.

\begin{figure*}
  \center
  \epsfig{width=0.49\textwidth, file=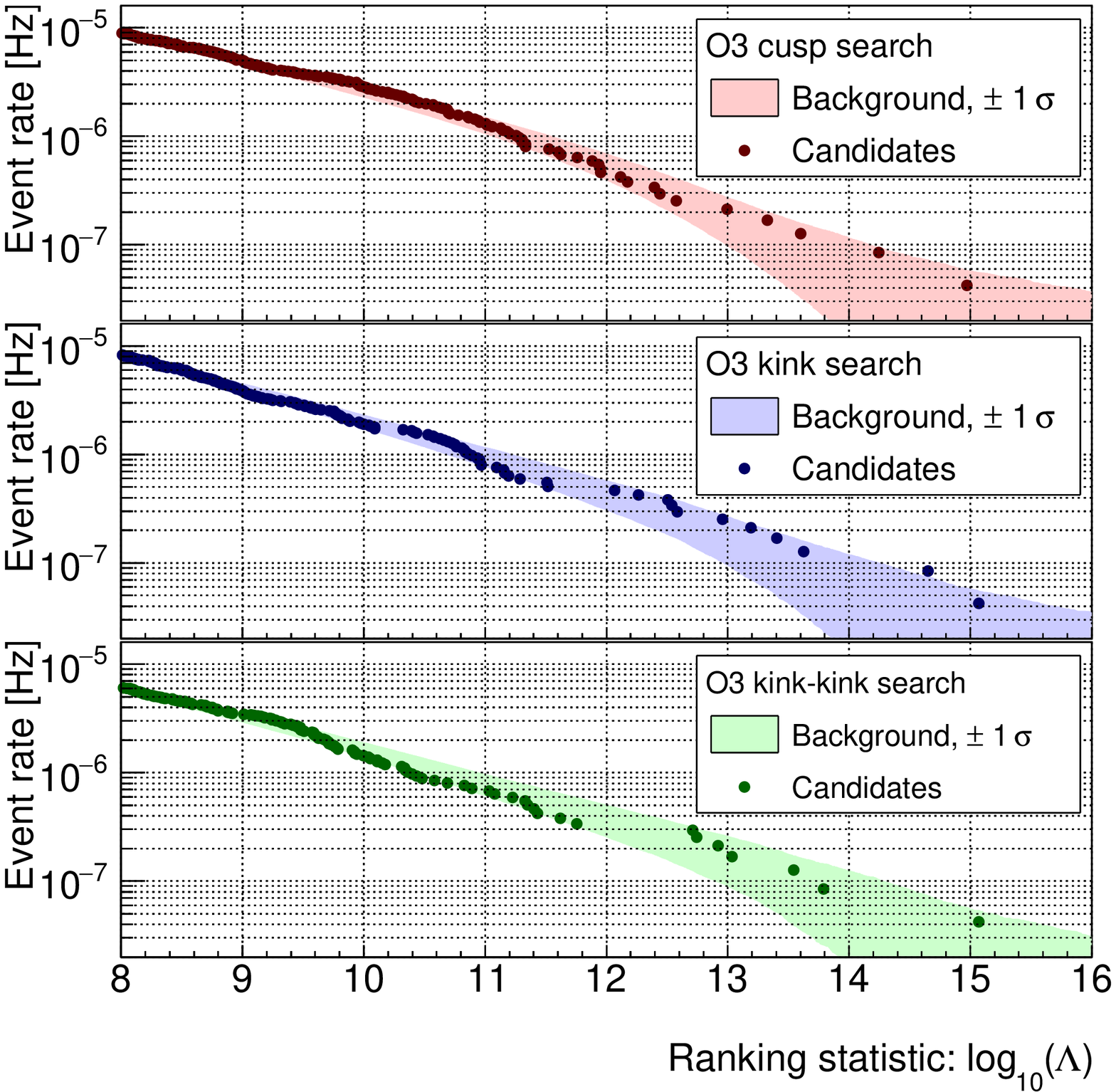}
  \epsfig{width=0.49\textwidth, file=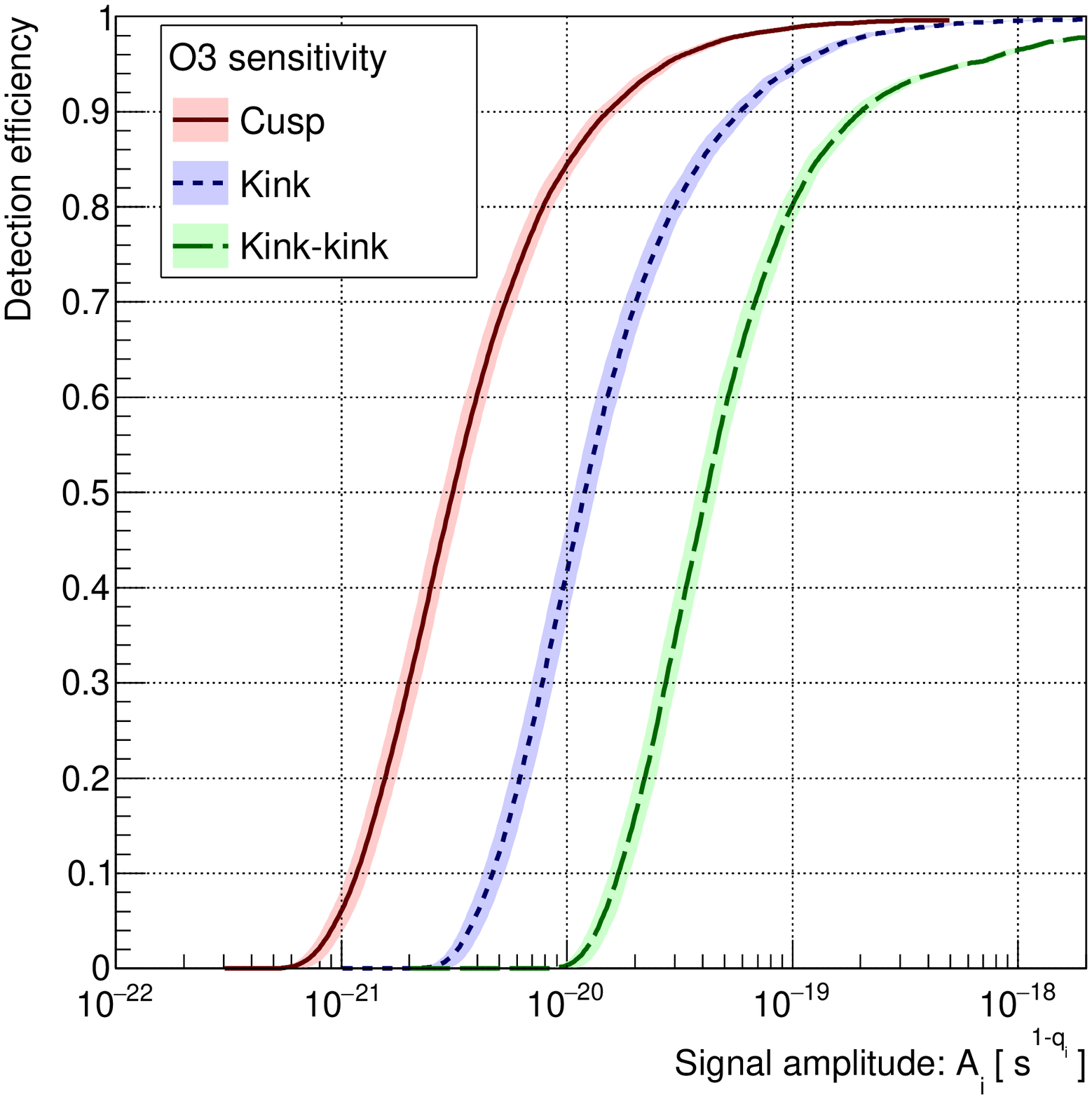}
  \caption{Left panel: cumulative distribution of cosmic string burst candidate events produced by cusps (top), kinks (middle) and kink-kink collisions (bottom). The expected distributions from background noise are represented by $\pm 1 \sigma$ shaded areas. Right panel: the detection efficiency is measured using simulated signals, as a function of the signal amplitude for cusps, kinks and kink-kink collisions. Note that the horizontal axis measures different amplitude quantities, $A_i$, for the three types of signals, parameterized by the waveform frequency power law $q_i$.}
  \label{fig:burst_results}
\end{figure*}

The left panel of Fig.~\ref{fig:burst_results} presents the cumulative distribution of coincident O3 burst events as a function of the likelihood ratio $\Lambda$ for the cusp, kink and kink-kink collision searches. To estimate the background noise associated with each search, time shifts are applied to each detector strain data such that no real gravitational-wave event can be found in coincidence. For this study, we use 300 time-shifts, totaling $T_{\mathrm{bkg}}=225$\,years of data containing only noise coincident events, the distribution of which is represented in the left panel of Fig.~\ref{fig:burst_results} with a $\pm 1\sigma$ shaded band. The candidate events, obtained with no time shift, are all compatible with the noise distribution within $\pm 2\sigma$. The cusp, kink and kink-kink collision waveforms are very similar, resulting in the loudest events being the same for the three searches. The ten loudest events were carefully scrutinized. They all originate from a well-known category of transient noise affecting all detectors, that are broadband and very short-duration noise events of unknown instrumental origin~\cite{TheLIGOScientific:2016zmo,Cabero:2019orq}.

From the non-detection result, we measure the LIGO--Virgo sensitivity to cosmic string signals by performing the burst search analysis over O3 data with injections of simulated cusp, kink and kink-collision waveforms. The amplitudes of injected signals comfortably cover the range where none to almost all the signals are detected. To recover injected signals, we use the loudest-event method described in~\cite{Brady:2004gt}, where the detection threshold is set to the level of the highest-ranked event found in the search: $\log_{10}(\Lambda)\simeq 15.0$, 15.1, and 15.1 for cusps, kinks, and kink-kink collisions, respectively. The resulting efficiencies $\varepsilon_i(A_i)$ as a function of the signal amplitude are presented in the right panel of Fig.~\ref{fig:burst_results}.
Cusp events directed at the Earth with amplitudes above $2\times 10^{-20} \mathrm{s}^{-1/3}$ can be detected at $\sim 90\%$. In terms of loop proper lengths, this corresponds, for example, to loops larger than $1.7 \times 10^{6} (G\mu/10^{-10})^{-3/2}$ light years at redshift 100.
The expected detection burst rate is calculated from the detection efficiency:
\begin{equation}
    R_i = \int \frac{\ud R_i}{\ud A_i}(A_i, f_*; G\mu, N_k) \varepsilon_i(A_i) ~\ud A_i.
\end{equation}
The detectable burst rate $\ud R_i/\ud A_i$ is obtained from Eq.~\ref{eq:drdz_gamma}, which can be expressed in terms of amplitude using Eq.~\ref{eq:Aq} and calculated for the lowest value of the high-frequency cutoff $f_*$ (see Supplemental Material for details).

We assume that the occurrence of a detectable burst of gravitational waves follows a Poisson distribution with mean given by the estimated detection rate.
For a set of parameters $(G\mu, N_{\rm k})$, models which predict a detection rate larger than $2.996 / T_\mathrm{obs}$ are excluded at $95\%$, i.e., we exclude models that predict a $>95\%$ confidence level detection.

%% file: Stoch_search.tex
\section{Stochastic Search} \label{sec:stoch_search}
A search for a stochastic gravitational wave background~\cite{Christensen:2018iqi} was carried out using the LIGO and Virgo O3 data~\cite{O3_isotropic} in which a correlated background in different interferometer pairs is sought. These results were combined with those from the previous two observing runs, O1 and O2~\cite{TheLIGOScientific:2016dpb,Abbott:2017mem,LIGOScientific:2019vic}. The results reported in~\cite{O3_isotropic} assume the normalized energy density of the stochastic background, Eq.~(\ref{sb}), to be a power-law $\alpha$ of the frequency:
\begin{equation}
  \Omega_{\rm GW}(f) = \Omega_{\rm ref} \left(\frac{f}{f_{\rm ref}}\right)^\alpha,
\end{equation}
where $f_{\rm ref}$ denotes a reference frequency, fixed to 25 Hz, a convenient choice in the sensitive part of the frequency band. LIGO and Virgo did not detect a stochastic background, and so set upper limits depending on the value of $\alpha$. The stochastic background from cosmic strings in the LIGO--Virgo frequency band is predicted to be approximately flat, setting the upper bound $\Omega_{\rm GW}\leq 5.8\times 10^{-9}$ at the $95\%$ credible level for a flat $\alpha = 0$ background and using a log-uniform prior in $\Omega_{\rm GW}$.

In the present study, we perform a Bayesian analysis taking into account the precise shape of the background (see Fig.~\ref{spec}) instead of a power-law and use it to derive upper limits on the cosmic string parameters.
We first calculate the log-likelihood function assuming a Gaussian distributed noise, which up to a constant is
\begin{equation}
  \ln \mathcal{L}(\hat C^{IJ}_a|G\mu,N_k) = - \frac{1}{2} \sum_{IJ,a}\frac{(\hat C^{IJ}_a-\Omega_{\rm GW}^{(M)}(f_a;G\mu,N_k))^2}{\sigma_{IJ}^2(f_a)} .
  \label{eq:likelihood}
\end{equation}
Here $\hat C^{IJ}_a \equiv \hat C^{IJ}(f_a)$, and $IJ$ are detector pairs L1-H1, L1-V1, and H1-V1.
$\hat C^{IJ}(f_a)$ and $\sigma^2(f_a)$ are, respectively, a cross-correlation estimator for the $IJ$ detecor pair and its variance at frequency $f_a$ as detailed in~\cite{Mandic_2012}. Following the same approach as in the O1 stochastic analysis we have used the frequency bins ranging from 20 to 86 Hz. The gravitational-wave energy density, $\Omega_{\rm GW}^{(M)}(f_a;G\mu,N_k)$, is predicted by the cosmic string model $M=\{{\bf{A}},{\bf{B}},{\bf{C}}\}$ and computed with Eq.~\ref{eq:Omega} at frequency $f_a$.

For our Bayesian analysis, we specify priors for the parameters in the cosmic string model, {\sl i.e.}, $p(G\mu | I_{G\mu})$ and $p(N_{\rm k} | I_{N_{\rm k}})$. The variables $I_{G\mu}$ and  $I_{N_{\rm k}}$ denote the information on the distributions of $G\mu$ and $N_k$, which are determined by theory predictions.  For $p(G \mu | I_{G\mu})$, we choose a log-uniform prior for $10^{-18}\le G\mu \le 10^{-6}$. Here the upper bound is set by the cosmic microwave background measurements \cite{Ade:2013xla,Lizarraga:2016onn,Henrot-Versille:2014jua,Lazanu:2014eya}. The lower bound is arbitrary, chosen for consistency with the study in \cite{Auclair:2019wcv}; we note, however, that our results remain almost unchanged if we choose a smaller value for the lower bound on $G\mu$. For $p(N_{\rm k} | I_{N_{\rm k}})$, we aim at constraining $G\mu$ for each choice of $N_{\rm k}$. Therefore the prior $p(N_{\rm k} | I_{N_{\rm k}})$ is taken to be a $\delta$-function for each value of $N_{\rm k}$. The number of kinks per loop oscillation $N_{\rm k}$ being fixed, the posterior for the parameter $G\mu$ is calculated according to Bayes' theorem:
\begin{eqnarray}
  p(G\mu | N_{\rm k}) \propto \mathcal{L}(\hat C^{IJ}_a|G\mu, N_{\rm k}) p(G\mu| I_{G\mu}) p(N_{\rm k} | I_{N_{\rm k}}) .
\end{eqnarray}
We calculate 95\% credible intervals for $G\mu$.

%% file: Constraints.tex
\begin{figure*}
  \centering
  \includegraphics[width=0.455\textwidth]{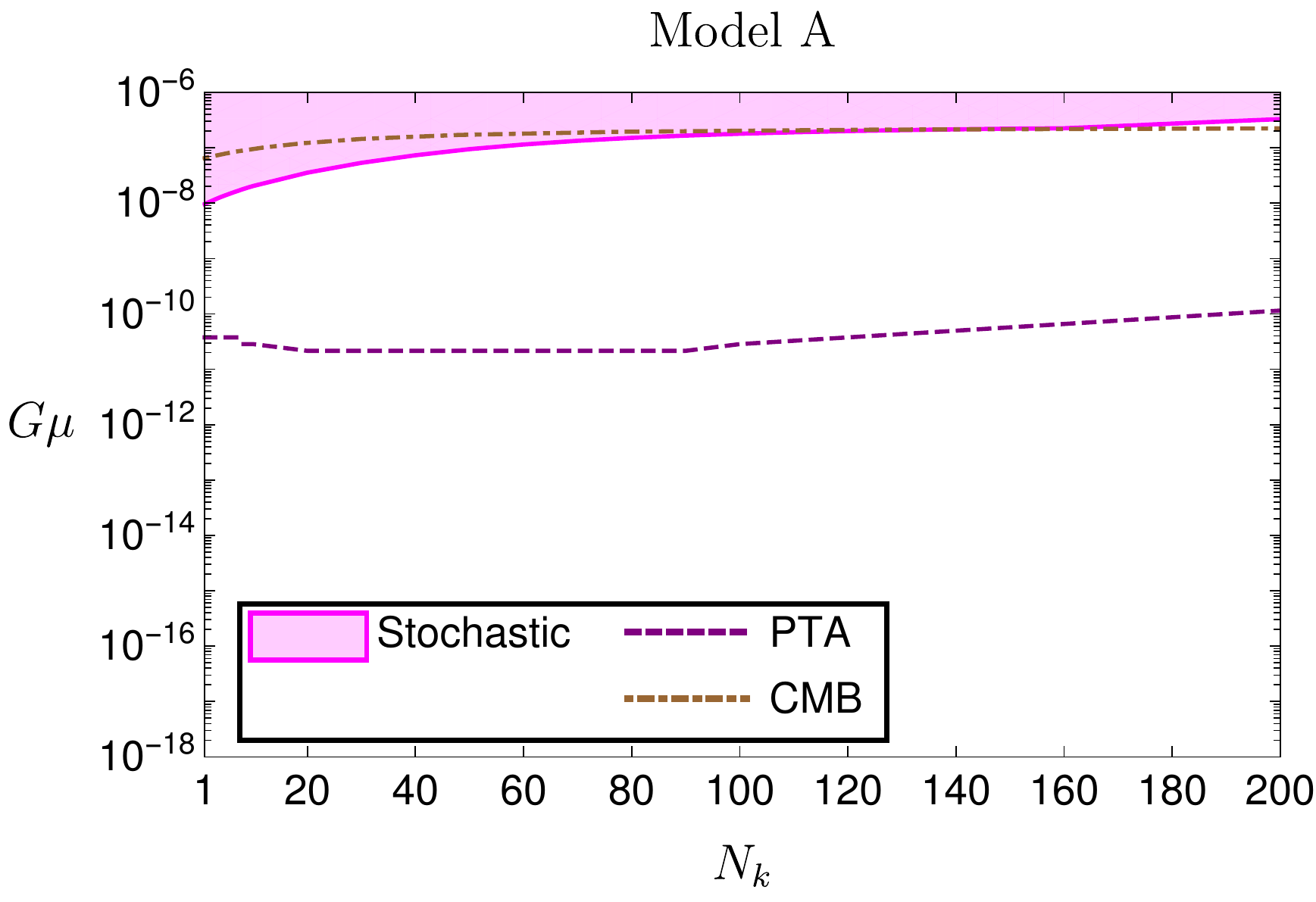}
  \includegraphics[width=0.455\textwidth]{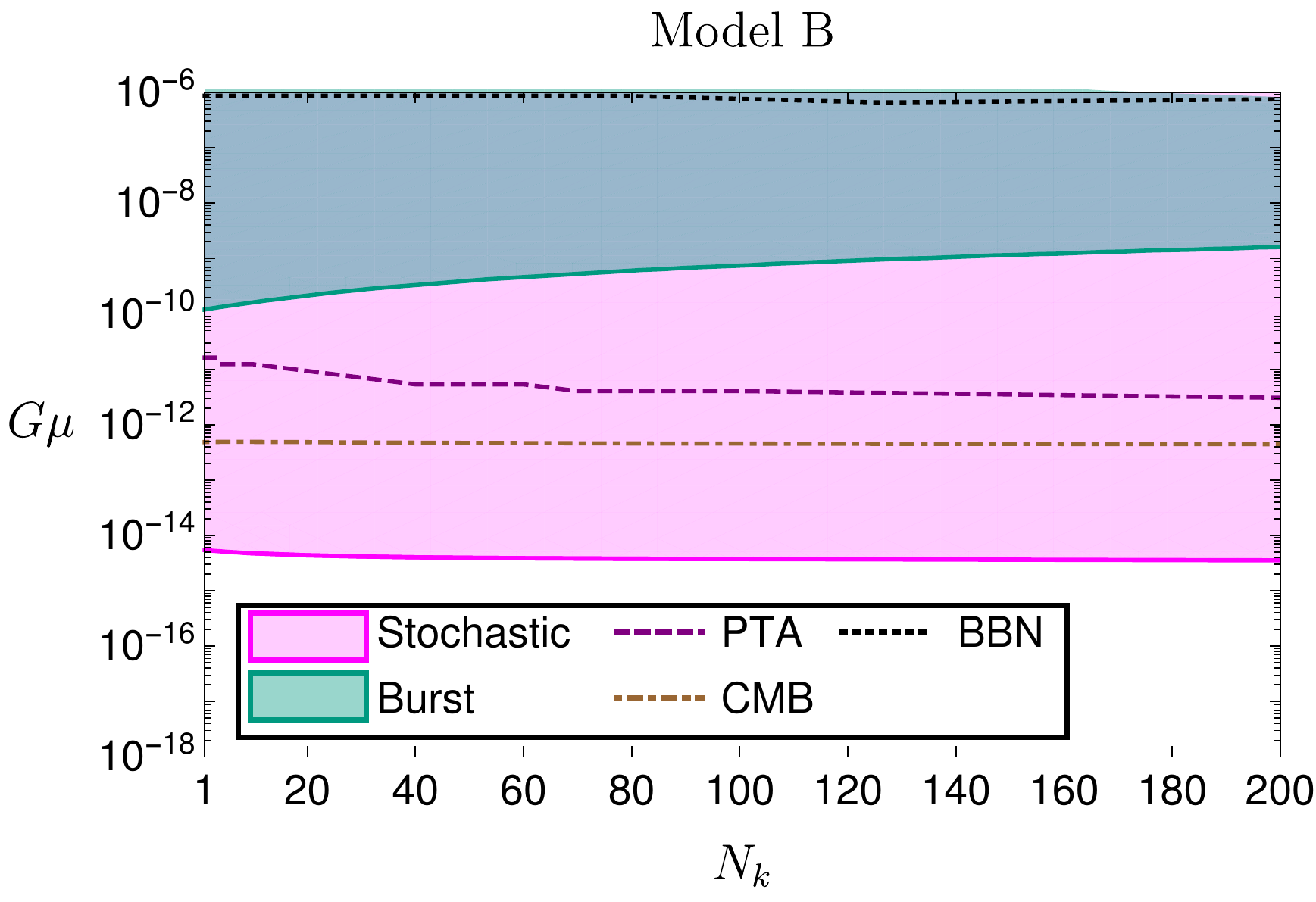}
  \\
  \includegraphics[width=0.455\textwidth]{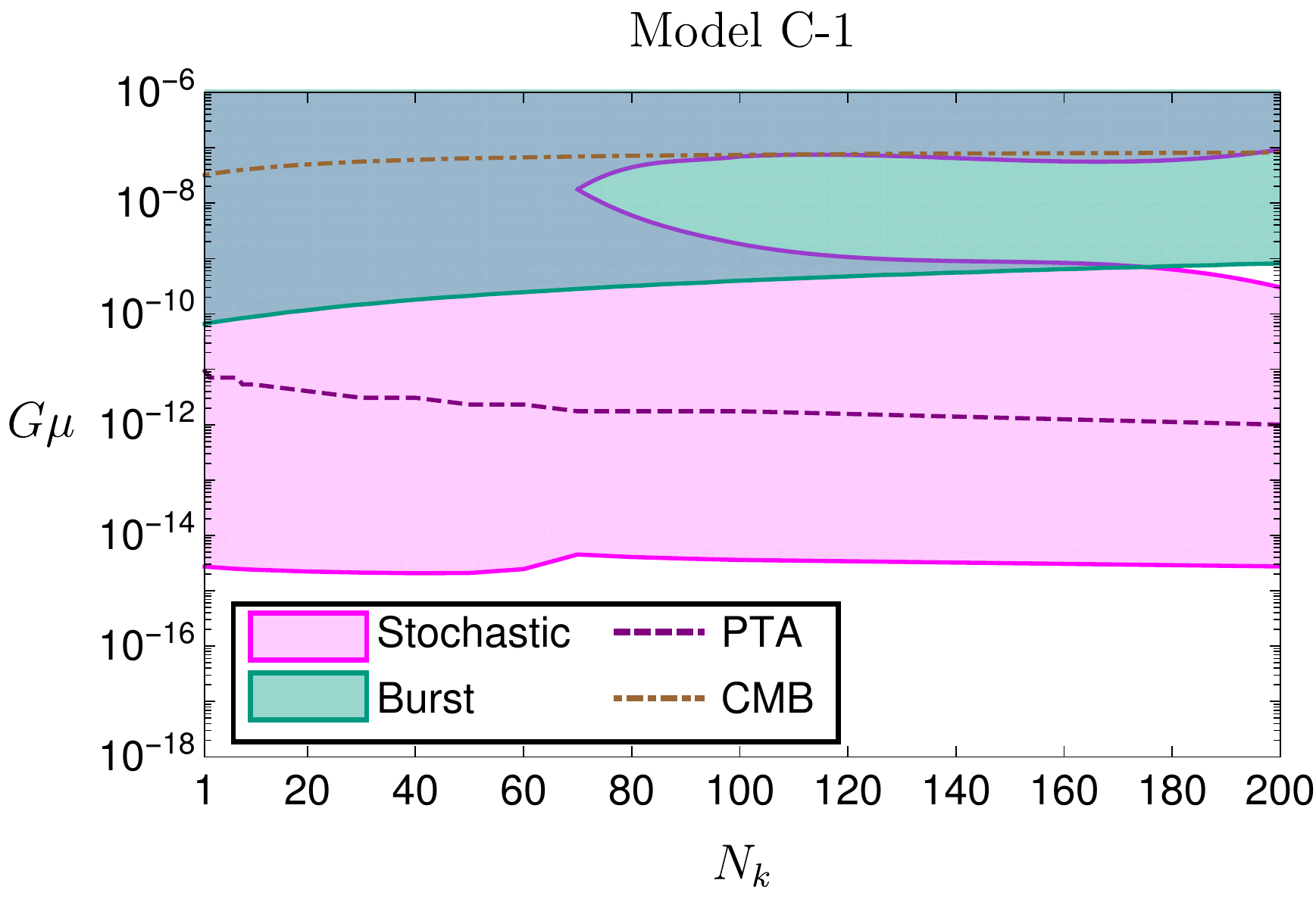}
  \includegraphics[width=0.455\textwidth]{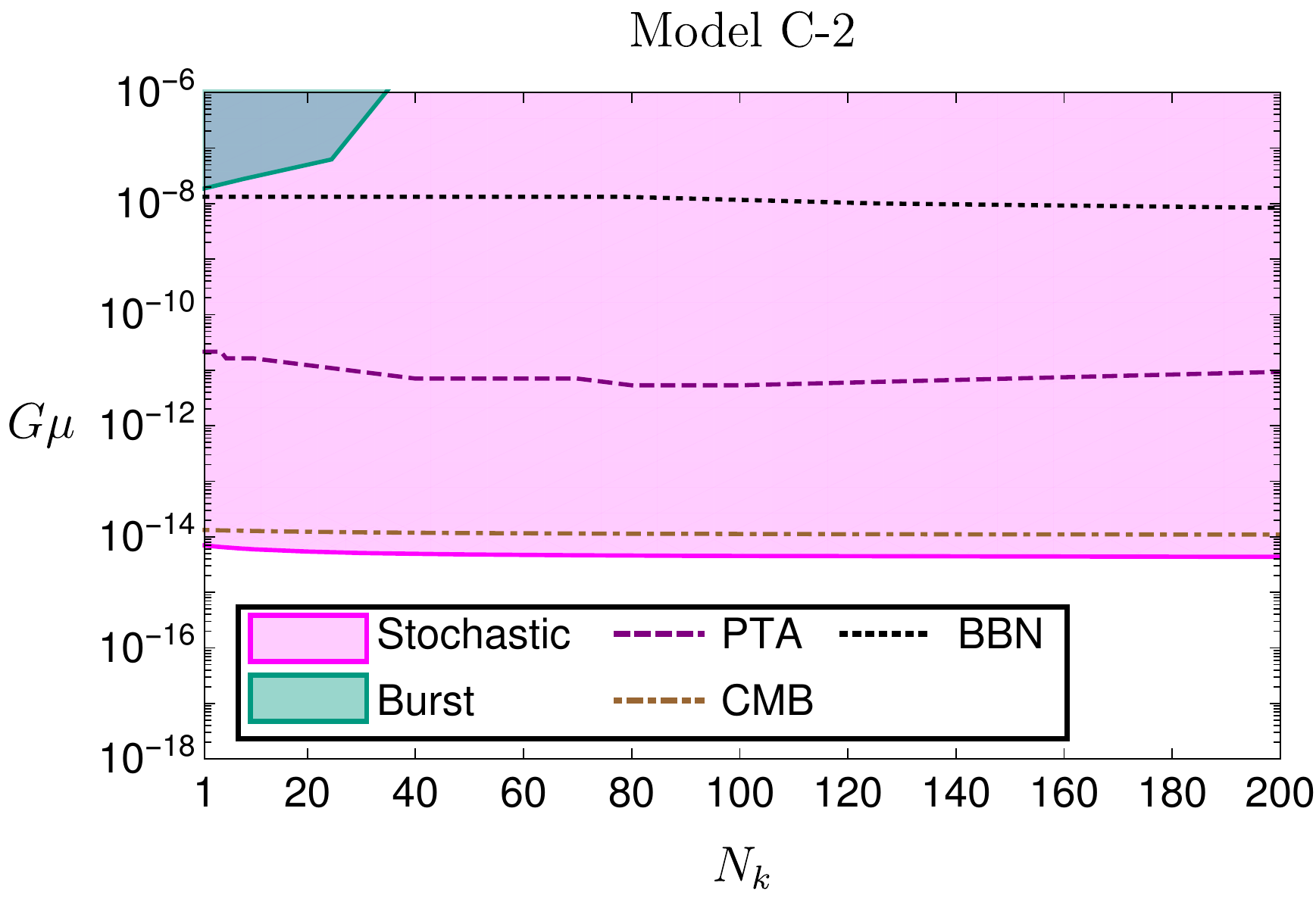}
  \caption{\label{fig:M23:constraints}
    Advanced LIGO--Virgo exclusion contours at 95\% C.L. on the cosmic string parameter space, $(N_{\rm k}, G\mu)$, derived from the stochastc search (pink), the burst search (turquoise) and both searches. Four models are considered to describe the distribution of cosmic string loops: model {\bf A} (top-left), model {\bf B} (top-right), model {\bf C}-1 (bottom-left) and model {\bf C}-2 (bottom-right). Note that the stochastic result combines the data of O1, O2 and O3 while the burst search only includes O3 data. We also report limits from other experiments: pulsar timing arrays (PTA), cosmic microwave background (CMB) and Big Bang nucleosynthesis~\cite{Lasky:2015lej}.
  }
  \label{SGWB-constr}
\end{figure*}

\section{Constraints}\label{sec:constraints}
We show in Fig.~\ref{SGWB-constr} the region of the $G\mu$ and $N_k$ parameter space excluded at the 95\% confidence level by the burst and stochastic searches; the number of cusps $N_{\rm c}$ being fixed to $1$.
For the stochastic search (Sec.~\ref{sec:stoch_search}) we present constraints from the combined O1+O2+O3 data; for the burst search (Sec.~\ref{sec:burst_search}) we derive constraints from the non-detection result using O3 data. We consider three models for the Nambu-Goto cosmic string loop distributions, dubbed {\bf A}, {\bf  B} and {\bf  C}. For the latter we choose two sets of benchmark numbers: for model \textbf{C}-1 we set $(\chir, \chim) = (0.45, 0.295)$ and for model \textbf{C}-2 $(\chir, \chim) = (0.2, 0.45)$ (see the Supplemental Material).

Using model {\bf A}, the derived gravitational-wave power spectrum is much weaker than in the other models, leading to weaker constraints. Model {\bf C}-2 mimics the loop production function of model \textbf{A} in the matter era and of model \textbf{B} in the radiation era. In the frequency band of LIGO--Virgo, the stochastic background is dominated by the contribution from loops in the radiation era, hence models  \textbf{B} and  \textbf{C}-2 give similar results. Conversely, the gravitational-wave power spectrum obtained from model {\bf C}-1, which mimics the loop production function of model \textbf{A} in the radiation era and of model \textbf{B} in the matter era, presents more subtle features. Larger values of $G\mu$ do not necessarily produce larger signal amplitudes, creating structures in the constraint plot. For an analytical understanding of these findings, we refer the reader to \cite{Auclair:2020oww}. For a better understanding of the loop visibility domain in terms of redshift, we refer to the Fig. 2 of \cite{Ringeval:2017eww}.

The stochastic analysis leads to the following constraints on $G\mu$. For model {\bf A}, we rule out the range
$G\mu\gtrsim (9.6 \times 10^{-9} - 10^{-6})$.
For model {\bf B}, we rule out:
$G\mu \gtrsim (4.0 - 6.3 ) \times 10^{-15} $. For model {\bf C}-1, we rule out
$G\mu \gtrsim (2.1 - 4.5) \times 10^{-15}$, aside from a small region where $N_{\rm k} \gtrsim 180 $. Finally, for model {\bf C}-2, we rule out:
$G\mu \gtrsim (4.2 - 7.0 ) \times 10^{-15}$. 

The burst search upper limits are not as stringent as the ones derived from the stochastic search. In particular, the constraints on the string tension for model {\bf A} are too weak to be represented in the figure. The only case where the burst analysis leads to tighter constraints, is for model {\bf C}-1 and  for $N_{\rm k}> 70$.

In the present analysis, the average number of cusps per oscillation on a loop has been set to $1$.
It has been shown that the number of cusps per period of string oscillation scales with the number of harmonics on the loop~\cite{Pazouli:2020qmj}.
Note that with many cusps on the string, the decay constant $\Gamma_{\rm d}$ is enhanced and the lifetime of the loop is hence greatly reduced.
Consequently, a high number of cusps on the loops gives qualitatively the same result as increasing the number of kinks: for model \textbf{A}, the constraints are weakened, whereas for models \textbf{B} and \textbf{C} the bounds are insensitive to $N_\mathrm{c}$; this has been confirmed by our numerical study.

One can also compare these results with limits obtained from pulsar timing array measurements, and indirect limits from Big Bang nucleosynthesis and cosmic microwave background data~\cite{Lasky:2015lej}. Repeating the analysis done in \cite{Abbott:2017mem} with $N_{\rm k}$ up to 200, we find that for model {\bf A}, the strongest limit comes from pulsar timing measurements, excluding string tensions $G\mu \gtrsim 10^{-10}$. For model {\bf B} and {\bf C}-1 the strongest limits are derived from the LIGO--Virgo stochastic search. Finally, for model {\bf C}-2, the cosmic microwave background constraint is almost as strong as the one obtained from the O1+O2+O3 stochastic search. The next observing run, O4, will give us a new opportunity to detect signals from cosmic strings.

%% file: Conclusions.tex
\section{Conclusions}
\label{sec:conclusions}

Using data from the third observing run of Advanced LIGO and Virgo, we have performed a burst and a stochastic gravitational wave background search to constrain the tension of Nambu-Goto strings, as a function of the number of kinks per oscillation, for four loop distributions.
We have tested models {\bf A} and {\bf B} already considered in the O1 and O2 analyses~\cite{LIGOScientific:2019vic}. The current constraints on $G\mu$ are stronger by two and one orders of magnitude for models {\bf A} and {\bf B}, respectively, when fixing $N_{\rm k}=1$. In addition, we have used two variants of a new model, dubbed model {\bf C}, that interpolates between models {\bf A} and {\bf B}. For the first time, we have studied the effect of kink-kink interactions, which is relevant for large numbers of kinks, and investigated the effect of a large number of cusps, as both effects are favored by cosmic string simulations. 


%% file: P2000488_v3.tex
The authors gratefully acknowledge the support of the United States
National Science Foundation (NSF) for the construction and operation of the
LIGO Laboratory and Advanced LIGO as well as the Science and Technology Facilities Council (STFC) of the
United Kingdom, the Max-Planck-Society (MPS), and the State of
Niedersachsen/Germany for support of the construction of Advanced LIGO 
and construction and operation of the GEO600 detector. 
Additional support for Advanced LIGO was provided by the Australian Research Council.
The authors gratefully acknowledge the Italian Istituto Nazionale di Fisica Nucleare (INFN),  
the French Centre National de la Recherche Scientifique (CNRS) and
the Netherlands Organization for Scientific Research, 
for the construction and operation of the Virgo detector
and the creation and support  of the EGO consortium. 
The authors also gratefully acknowledge research support from these agencies as well as by 
the Council of Scientific and Industrial Research of India, 
the Department of Science and Technology, India,
the Science \& Engineering Research Board (SERB), India,
the Ministry of Human Resource Development, India,
the Spanish Agencia Estatal de Investigaci\'on,
the Vicepresid\`encia i Conselleria d'Innovaci\'o, Recerca i Turisme and the Conselleria d'Educaci\'o i Universitat del Govern de les Illes Balears,
the Conselleria d'Innovaci\'o, Universitats, Ci\`encia i Societat Digital de la Generalitat Valenciana and
the CERCA Programme Generalitat de Catalunya, Spain,
the National Science Centre of Poland and the Foundation for Polish Science (FNP),
the Swiss National Science Foundation (SNSF),
the Russian Foundation for Basic Research, 
the Russian Science Foundation,
the European Commission,
the European Regional Development Funds (ERDF),
the Royal Society, 
the Scottish Funding Council, 
the Scottish Universities Physics Alliance, 
the Hungarian Scientific Research Fund (OTKA),
the French Lyon Institute of Origins (LIO),
the Belgian Fonds de la Recherche Scientifique (FRS-FNRS), 
Actions de Recherche Concertées (ARC) and
Fonds Wetenschappelijk Onderzoek – Vlaanderen (FWO), Belgium,
the Paris \^{I}le-de-France Region, 
the National Research, Development and Innovation Office Hungary (NKFIH), 
the National Research Foundation of Korea,
the Natural Science and Engineering Research Council Canada,
Canadian Foundation for Innovation (CFI),
the Brazilian Ministry of Science, Technology, and Innovations,
the International Center for Theoretical Physics South American Institute for Fundamental Research (ICTP-SAIFR), 
the Research Grants Council of Hong Kong,
the National Natural Science Foundation of China (NSFC),
the Leverhulme Trust, 
the Research Corporation, 
the Ministry of Science and Technology (MOST), Taiwan,
the United States Department of Energy,
and
the Kavli Foundation.
The authors gratefully acknowledge the support of the NSF, STFC, INFN and CNRS for provision of computational resources.
This work was supported by MEXT, JSPS Leading-edge Research Infrastructure Program, JSPS Grant-in-Aid for Specially Promoted Research 26000005, JSPS Grant-in-Aid for Scientific Research on Innovative Areas 2905: JP17H06358, JP17H06361 and JP17H06364, JSPS Core-to-Core Program A. Advanced Research Networks, JSPS Grant-in-Aid for Scientific Research (S) 17H06133, the joint research program of the Institute for Cosmic Ray Research, University of Tokyo, National Research Foundation (NRF) and Computing Infrastructure Project of KISTI-GSDC in Korea, Academia Sinica (AS), AS Grid Center (ASGC) and the Ministry of Science and Technology (MoST) in Taiwan under grants including AS-CDA-105-M06, Advanced Technology Center (ATC) of NAOJ, and Mechanical Engineering Center of KEK.

%% file: Appendix_model.tex
\section{Loop distributions}\label{appx:loop}

For model {\bf A}, the loop distribution is given by the sum of
\begin{multline}
    t^4 \n_{\rm rad} = \frac{0.18\times 2\sqrt{\uH}\orad^{3/4}}{(\gamma + \Gamma_{\rm d} G\mu)^{5/2}}  t^{3/2} (1+z)^3 \\
    \times \begin{cases}
        \Theta\left(0.18 t - \ell\right) & \text{ if } t<t_{\rm eq} \\
        \Theta \left[0.18 t_{\rm eq} - \ell + \Gamma_{\rm d} G\mu(t-t_{\rm eq})\right] & \text{ if } t>t_{\rm eq}
    \end{cases}
\end{multline}
for loops produced during the radiation era, and
\begin{multline}
    t^4 \n_{\rm mat}= \frac{0.27 - 0.45\gamma^{0.31}}{(\gamma + \Gamma_{\rm d} G\mu)^2} \Theta(0.18 - \gamma) \\ \times \Theta \left[\ell + \Gamma_{\rm d} G\mu(t-t_{\rm eq}) - 0.18 t_{\rm eq}\right]
\end{multline}
for loops produced during the matter era. Note that $t_{\rm eq}$ is the time of the radiation to matter transition.

For model {\bf B}, the loop distribution is the sum of
\begin{equation}
    t^4 \n_{\rm rad} =
    \begin{cases}
        \dfrac{0.08}{(\gamma + \Gamma_{\rm d} G\mu)^{3-2\chir}} & \text{ if } \Gamma_{\rm d} G\mu < \gamma\\
        \dfrac{0.08 (1/2 - 2\chir)}{(2-2\chir) \Gamma_{\rm d} G\mu \gamma^{2-2\chir}} & \text{ if } \gamma_{\rm c} < \gamma < \Gamma_{\rm d} G\mu\\
        \dfrac{0.08(1/2 - 2\chir)}{(2-2\chir) \Gamma_{\rm d} G\mu \gamma_c^{2-2\chir}} & \text{ if } \gamma < \gamma_{\rm c}\\
    \end{cases}
\end{equation}
for loops produced during the radiation era,
\begin{equation}
    t^4 \n_{\rm mat} =
    \begin{cases}
        \dfrac{0.015}{(\gamma + \Gamma_{\rm d} G\mu)^{3-2\chim}} & \text{ if } \Gamma_{\rm d} G\mu < \gamma\\
        \dfrac{0.015 (1 - 2\chim)}{(2-2\chim) \Gamma_{\rm d} G\mu \gamma^{2-2\chim}} & \text{ if } \gamma_{\rm c} < \gamma < \Gamma_{\rm d} G\mu\\
        \dfrac{0.015(1 - 2\chim)}{(2-2\chim) \Gamma_{\rm d} G\mu \gamma_c^{2-2\chim}} & \text{ if } \gamma < \gamma_{\rm c}\\
    \end{cases}
\end{equation}
for loops produced during the matter era, and
\begin{multline}
    t^4 \n_{\rm rad-mat}= \left(\frac{t}{t_{\rm eq}}\right)^3 \left(\frac{1+z}{1+z_{\rm eq}}\right)^3 t_{\rm eq}^4 \\
    \times \n_{\rm rad}^{(3)}\left[\frac{\gamma t + \Gamma_{\rm d} G\mu (t-t_{\rm eq})}{t_{\rm eq}}\right]
\end{multline}
for loops produced during the radiation era and decaying during the matter era. The subscripts $_{\rm rad}$ and $_{\rm mat}$ refer to the radiation- and matter-dominated eras, respectively, and $\gamma_{\rm c}$ is the gravitational back-reaction scale.

For model {\bf C}, the loop distribution can be approximated in the radiation era as
\begin{multline}
    t^4 \n_{\rm rad} = \frac{c_{\rm rad}}{1/2 - 2\chir} \\
    \times \begin{cases}
        (\gamma + \Gamma_{\rm d} G\mu)^{2\chir - 3} -  \dfrac{\gamma_\infty^{2\chir-1/2}}{(\gamma + \Gamma_\mathrm{d} G\mu) ^ {5/2}} & \text{ if } \Gamma_{\rm d} G\mu < \gamma\\
        \dfrac{\gamma^{2\chir - 2}}{(2-2\chir) \Gamma_{\rm d} G\mu } -  \dfrac{\gamma_\infty^{2\chir-1/2}}{(\gamma + \Gamma_\mathrm{d} G\mu) ^ {5/2}} & \text{ if } \gamma_{\rm c} < \gamma < \Gamma_{\rm d} G\mu\\
        \dfrac{\gamma_{\rm c}^{2\chir - 2}}{(2-2\chir) \Gamma_{\rm d} G\mu } -  \dfrac{\gamma_\infty^{2\chir-1/2}}{(\gamma + \Gamma_\mathrm{d} G\mu) ^ {5/2}} & \text{ if } \gamma < \gamma_{\rm c}\\
    \end{cases}
\end{multline}
and in the matter era as
\begin{multline}
    t^4 \n_{\rm mat} = \frac{c_{\rm mat}}{1 - 2\chim} \\
    \times \begin{cases}
        (\gamma + \Gamma_{\rm d} G\mu)^{2\chim - 3} -  \dfrac{\gamma_\infty^{2\chim-1}}{(\gamma + \Gamma_\mathrm{d} G\mu) ^ {2}} & \text{ if } \Gamma_{\rm d} G\mu < \gamma\\
        \dfrac{\gamma^{2\chim - 2}}{(2-2\chim) \Gamma_{\rm d} G\mu } -  \dfrac{\gamma_\infty^{2\chim-1}}{(\gamma + \Gamma_\mathrm{d} G\mu) ^ {2}} & \text{ if } \gamma_c < \gamma < \Gamma_{\rm d} G\mu\\
        \dfrac{\gamma_{\rm c}^{2\chim - 2}}{(2-2\chim) \Gamma_{\rm d} G\mu } -  \dfrac{\gamma_\infty^{2\chim-1}}{(\gamma + \Gamma_\mathrm{d} G\mu) ^ {2}} & \text{ if } \gamma < \gamma_c\\
    \end{cases}
\end{multline}
where $\gamma_\infty$ is the size of the largest loops in scaling units.

%% file: Appendix_burst.tex
\section{The burst analysis pipeline}\label{appx:burst}

The cosmic string burst search pipeline is divided into three main analysis steps. First, the cosmic string gravitational waveform is searched in LIGO and Virgo data using match-filtering techniques. A bank of waveform templates is chosen to match the expected signal in the frequency domain:
\begin{equation}
  t_{i,j}(f) = Af^{-q_i}\Theta(f_j-f)\Theta(f-f_l).
  \label{eq:burst_signal}
\end{equation}
The spectral index $q_i$ is taken from Eq.~\ref{eq:h} for cusps, kinks or kink-kink collisions. The waveform frequency $f$ is limited in range using the Heaviside function $\Theta$. The low-frequency cut-off $f_l$, resulting from the size of the feature producing the gravitational waves, takes values well below the sensitive band of the LIGO and Virgo detectors. We take $f_l=16$\,Hz. The high-frequency cutoff $f_j$ is a consequence of the gravitational-wave emission being beamed in a cone with an opening angle $\theta_m$, given in Eq.~\ref{eq:thetam}. The angle between the line of sight and the gravitational-wave direction, $\theta$, must be smaller than $\theta_m/2$, yielding $f<[2g_2\ell(1+z)\theta^3]^{-1}$. This high-frequency cutoff is unknown and is considered as free parameter, taking discrete values indexed by $j$. When searching for gravitational waves produced by cusps and kinks, we use 31 templates with high-frequency cutoff values distributed between 30\,Hz and 4096\,Hz and spaced in such a way that we guarantee less than 0.1\,\% loss in the signal-to-noise ratio due to template mismatch. For kink-kink collisions, the gravitational-wave radiation is isotropic and a single template covering the entire frequency band is used. The template amplitude $A$ is scaled when performing the match-filtering to measure the signal-to-noise ratio $\rho_{i,j}$ for each template $j$ tested with the LIGO and Virgo respective strain data $h_{\mathrm{det}}$:
\begin{equation}
  \rho_{i,j,\mathrm{det}}(t) = 4\Re\int_0^{\infty}{\frac{t^{*}_{i,j}(f)h_{\mathrm{det}}(f)}{S_n(f)}\exp(2\sqrt{-1}\pi ft)\ud f}.
\end{equation}
Here, $S_n(f)$ is the single-sided noise power spectral density of the detector. It is estimated locally over a few minutes of detector data.

The O3 LIGO--Virgo dataset is analyzed in five consecutive chunks of data to account for the detectors' noise evolution over the entire run. The chunk boundaries are defined by sudden changes of sensitivity of one detector and by commissioning interventions, including the 1-month commissioning break in October 2019. The signal-to-noise ratio time-series $\rho_{i,j,\mathrm{det}}(t)$ is computed for each detector, accumulating a total of 245 days, 252 days and 250 days for the LIGO--Livingston, the LIGO--Hanford and the Virgo detector respectively. The signal-to-noise ratio time-series are required to be above 3.75 and are clustered among templates. We obtain a set of events, or triggers, the parameters of which are given by the template with the highest signal-to-noise ratio peak value.

After the match-filtering step, a time coincidence is performed pair-wise between the triggers of each detector using a time window wide enough to account for the maximum light-travel time between detectors and calibration time uncertainties. A resulting set of double- and triple-coincident events is obtained when at least two detectors are taking data in nominal conditions, yielding a total observation time of $T_{\mathrm{obs}}=273.5$~days.

Finally, to discriminate astrophysical signals from background noise, we apply the multivariate method described in~\cite{Cannon:2008zz}, which uses a set of simulated cosmic string events and typical noise events to statistically infer the probability for a coincident event to be signal or noise. Hence, a likelihood ratio, $\Lambda$, is constructed with parameters characterizing the event~\cite{Aasi:2013vna}. For the O3 analysis, we introduce for the first time the event duration as a new discriminating variable. The event duration is defined as the duration for which $\rho_{i,j,\mathrm{det}}(t)$ remains above threshold considering all templates $j$ and using a tolerance of 100\,ms. Using this parameter, coupled to the signal-to-noise ratio, allows us to reject a large population of long-duration and low signal-to-noise ratio transient noise events contributing to the search background.

%% file: LSC-Virgo-KAGRA-Authors-Aug-2020-prd.tex
\author{R.~Abbott}
\affiliation{LIGO Laboratory, California Institute of Technology, Pasadena, CA 91125, USA}
\author{T.~D.~Abbott}
\affiliation{Louisiana State University, Baton Rouge, LA 70803, USA}
\author{S.~Abraham}
\affiliation{Inter-University Centre for Astronomy and Astrophysics, Pune 411007, India}
\author{F.~Acernese}
\affiliation{Dipartimento di Farmacia, Universit\`a di Salerno, I-84084 Fisciano, Salerno, Italy  }
\affiliation{INFN, Sezione di Napoli, Complesso Universitario di Monte S.Angelo, I-80126 Napoli, Italy  }
\author{K.~Ackley}
\affiliation{OzGrav, School of Physics \& Astronomy, Monash University, Clayton 3800, Victoria, Australia}
\author{A.~Adams}
\affiliation{Christopher Newport University, Newport News, VA 23606, USA}
\author{C.~Adams}
\affiliation{LIGO Livingston Observatory, Livingston, LA 70754, USA}
\author{R.~X.~Adhikari}
\affiliation{LIGO Laboratory, California Institute of Technology, Pasadena, CA 91125, USA}
\author{V.~B.~Adya}
\affiliation{OzGrav, Australian National University, Canberra, Australian Capital Territory 0200, Australia}
\author{C.~Affeldt}
\affiliation{Max Planck Institute for Gravitational Physics (Albert Einstein Institute), D-30167 Hannover, Germany}
\affiliation{Leibniz Universit\"at Hannover, D-30167 Hannover, Germany}
% ...

\collaboration{The LIGO Scientific Collaboration, the Virgo Collaboration, and the KAGRA Collaboration}

%% file: main.bbl
\begin{thebibliography}{64}%
\makeatletter
\providecommand \@ifxundefined [1]{%
 \@ifx{#1\undefined}
}%
\providecommand \@ifnum [1]{%
 \ifnum #1\expandafter \@firstoftwo
 \else \expandafter \@secondoftwo
 \fi
}%
\providecommand \@ifx [1]{%
 \ifx #1\expandafter \@firstoftwo
 \else \expandafter \@secondoftwo
 \fi
}%
\providecommand \natexlab [1]{#1}%
\providecommand \enquote  [1]{``#1''}%
\providecommand \bibnamefont  [1]{#1}%
\providecommand \bibfnamefont [1]{#1}%
\providecommand \citenamefont [1]{#1}%
\providecommand \href@noop [0]{\@secondoftwo}%
\providecommand \href [0]{\begingroup \@sanitize@url \@href}%
\providecommand \@href[1]{\@@startlink{#1}\@@href}%
\providecommand \@@href[1]{\endgroup#1\@@endlink}%
\providecommand \@sanitize@url [0]{\catcode `\\12\catcode `\$12\catcode
  `\&12\catcode `\#12\catcode `\^12\catcode `\_12\catcode `\%12\relax}%
\providecommand \@@startlink[1]{}%
\providecommand \@@endlink[0]{}%
\providecommand \url  [0]{\begingroup\@sanitize@url \@url }%
\providecommand \@url [1]{\endgroup\@href {#1}{\urlprefix }}%
\providecommand \urlprefix  [0]{URL }%
\providecommand \Eprint [0]{\href }%
\providecommand \doibase [0]{http://dx.doi.org/}%
\providecommand \selectlanguage [0]{\@gobble}%
\providecommand \bibinfo  [0]{\@secondoftwo}%
\providecommand \bibfield  [0]{\@secondoftwo}%
\providecommand \translation [1]{[#1]}%
\providecommand \BibitemOpen [0]{}%
\providecommand \bibitemStop [0]{}%
\providecommand \bibitemNoStop [0]{.\EOS\space}%
\providecommand \EOS [0]{\spacefactor3000\relax}%
\providecommand \BibitemShut  [1]{\csname bibitem#1\endcsname}%
\let\auto@bib@innerbib\@empty
\bibitem [{\citenamefont {Aasi}\ \emph {et~al.}(2015)\citenamefont {Aasi} \emph
  {et~al.}}]{0264-9381-32-7-074001}%
  \BibitemOpen
  \bibfield  {author} {\bibinfo {author} {\bibfnamefont {J.}~\bibnamefont
  {Aasi}} \emph {et~al.},\ }\href@noop {} {\bibfield  {journal} {\bibinfo
  {journal} {Classical and Quantum Gravity}\ }\textbf {\bibinfo {volume}
  {32}},\ \bibinfo {pages} {074001} (\bibinfo {year} {2015})}\BibitemShut
  {NoStop}%
\bibitem [{\citenamefont {Acernese}\ \emph {et~al.}(2015)\citenamefont
  {Acernese} \emph {et~al.}}]{0264-9381-32-2-024001}%
  \BibitemOpen
  \bibfield  {author} {\bibinfo {author} {\bibfnamefont {F.}~\bibnamefont
  {Acernese}} \emph {et~al.},\ }\href@noop {} {\bibfield  {journal} {\bibinfo
  {journal} {Classical and Quantum Gravity}\ }\textbf {\bibinfo {volume}
  {32}},\ \bibinfo {pages} {024001} (\bibinfo {year} {2015})}\BibitemShut
  {NoStop}%
\bibitem [{\citenamefont {Abbott}\ \emph
  {et~al.}(2020{\natexlab{a}})\citenamefont {Abbott} \emph
  {et~al.}}]{Abbott:2020niy}%
  \BibitemOpen
  \bibfield  {author} {\bibinfo {author} {\bibfnamefont {R.}~\bibnamefont
  {Abbott}} \emph {et~al.} (\bibinfo {collaboration} {LIGO Scientific,
  Virgo}),\ }\href@noop {} {\  (\bibinfo {year} {2020}{\natexlab{a}})},\
  \Eprint {http://arxiv.org/abs/2010.14527} {arXiv:2010.14527 [gr-qc]}
  \BibitemShut {NoStop}%
\bibitem [{\citenamefont {Abbott}\ \emph
  {et~al.}(2020{\natexlab{b}})\citenamefont {Abbott} \emph
  {et~al.}}]{Abbott:2020gyp}%
  \BibitemOpen
  \bibfield  {author} {\bibinfo {author} {\bibfnamefont {R.}~\bibnamefont
  {Abbott}} \emph {et~al.} (\bibinfo {collaboration} {LIGO Scientific,
  Virgo}),\ }\href@noop {} {\  (\bibinfo {year} {2020}{\natexlab{b}})},\
  \Eprint {http://arxiv.org/abs/2010.14533} {arXiv:2010.14533 [astro-ph.HE]}
  \BibitemShut {NoStop}%
\bibitem [{\citenamefont {Riles}(2013)}]{Riles20131}%
  \BibitemOpen
  \bibfield  {author} {\bibinfo {author} {\bibfnamefont {K.}~\bibnamefont
  {Riles}},\ }\href {\doibase http://dx.doi.org/10.1016/j.ppnp.2012.08.001}
  {\bibfield  {journal} {\bibinfo  {journal} {Progress in Particle and Nuclear
  Physics}\ }\textbf {\bibinfo {volume} {68}},\ \bibinfo {pages} {1 } (\bibinfo
  {year} {2013})}\BibitemShut {NoStop}%
\bibitem [{\citenamefont {Kibble}(1976)}]{Kibble:1976sj}%
  \BibitemOpen
  \bibfield  {author} {\bibinfo {author} {\bibfnamefont {T.~W.~B.}\
  \bibnamefont {Kibble}},\ }\href {\doibase 10.1088/0305-4470/9/8/029}
  {\bibfield  {journal} {\bibinfo  {journal} {J.Phys.A}\ }\textbf {\bibinfo
  {volume} {A 9}},\ \bibinfo {pages} {1387} (\bibinfo {year}
  {1976})}\BibitemShut {NoStop}%
\bibitem [{\citenamefont {Vilenkin}\ and\ \citenamefont
  {Shellard}(2000)}]{Vilenkin:2000jqa}%
  \BibitemOpen
  \bibfield  {author} {\bibinfo {author} {\bibfnamefont {A.}~\bibnamefont
  {Vilenkin}}\ and\ \bibinfo {author} {\bibfnamefont {E.~P.~S.}\ \bibnamefont
  {Shellard}},\ }\href@noop {} {\emph {\bibinfo {title} {{Cosmic Strings and
  Other Topological Defects}}}}\ (\bibinfo  {publisher} {Cambridge University
  Press},\ \bibinfo {year} {2000})\BibitemShut {NoStop}%
\bibitem [{\citenamefont {Hindmarsh}\ and\ \citenamefont
  {Kibble}(1995)}]{Hindmarsh:1994re}%
  \BibitemOpen
  \bibfield  {author} {\bibinfo {author} {\bibfnamefont {M.~B.}\ \bibnamefont
  {Hindmarsh}}\ and\ \bibinfo {author} {\bibfnamefont {T.~W.~B.}\ \bibnamefont
  {Kibble}},\ }\href {\doibase 10.1088/0034-4885/58/5/001} {\bibfield
  {journal} {\bibinfo  {journal} {Rept. Prog. Phys.}\ }\textbf {\bibinfo
  {volume} {58}},\ \bibinfo {pages} {477} (\bibinfo {year} {1995})},\ \Eprint
  {http://arxiv.org/abs/hep-ph/9411342} {arXiv:hep-ph/9411342 [hep-ph]}
  \BibitemShut {NoStop}%
\bibitem [{\citenamefont {Vachaspati}\ \emph {et~al.}(2015)\citenamefont
  {Vachaspati}, \citenamefont {Pogosian},\ and\ \citenamefont
  {Steer}}]{Vachaspati:2015cma}%
  \BibitemOpen
  \bibfield  {author} {\bibinfo {author} {\bibfnamefont {T.}~\bibnamefont
  {Vachaspati}}, \bibinfo {author} {\bibfnamefont {L.}~\bibnamefont
  {Pogosian}}, \ and\ \bibinfo {author} {\bibfnamefont {D.}~\bibnamefont
  {Steer}},\ }\href {\doibase 10.4249/scholarpedia.31682} {\bibfield  {journal}
  {\bibinfo  {journal} {Scholarpedia}\ }\textbf {\bibinfo {volume} {10}},\
  \bibinfo {pages} {31682} (\bibinfo {year} {2015})},\ \Eprint
  {http://arxiv.org/abs/1506.04039} {arXiv:1506.04039 [astro-ph.CO]}
  \BibitemShut {NoStop}%
\bibitem [{\citenamefont {Jeannerot}\ \emph {et~al.}(2003)\citenamefont
  {Jeannerot}, \citenamefont {Rocher},\ and\ \citenamefont
  {Sakellariadou}}]{Jeannerot:2003qv}%
  \BibitemOpen
  \bibfield  {author} {\bibinfo {author} {\bibfnamefont {R.}~\bibnamefont
  {Jeannerot}}, \bibinfo {author} {\bibfnamefont {J.}~\bibnamefont {Rocher}}, \
  and\ \bibinfo {author} {\bibfnamefont {M.}~\bibnamefont {Sakellariadou}},\
  }\href {\doibase 10.1103/PhysRevD.68.103514} {\bibfield  {journal} {\bibinfo
  {journal} {Phys.Rev.}\ }\textbf {\bibinfo {volume} {D 68}},\ \bibinfo {pages}
  {103514} (\bibinfo {year} {2003})},\ \Eprint
  {http://arxiv.org/abs/hep-ph/0308134} {arXiv:hep-ph/0308134 [hep-ph]}
  \BibitemShut {NoStop}%
\bibitem [{\citenamefont {Vachaspati}\ and\ \citenamefont
  {Vilenkin}(1985)}]{Vachaspati:1984gt}%
  \BibitemOpen
  \bibfield  {author} {\bibinfo {author} {\bibfnamefont {T.}~\bibnamefont
  {Vachaspati}}\ and\ \bibinfo {author} {\bibfnamefont {A.}~\bibnamefont
  {Vilenkin}},\ }\href {\doibase 10.1103/PhysRevD.31.3052} {\bibfield
  {journal} {\bibinfo  {journal} {Phys. Rev.}\ }\textbf {\bibinfo {volume}
  {D31}},\ \bibinfo {pages} {3052} (\bibinfo {year} {1985})}\BibitemShut
  {NoStop}%
\bibitem [{\citenamefont {Sakellariadou}(1990)}]{Sakellariadou:1990ne}%
  \BibitemOpen
  \bibfield  {author} {\bibinfo {author} {\bibfnamefont {M.}~\bibnamefont
  {Sakellariadou}},\ }\href {\doibase 10.1103/PhysRevD.42.354,
  10.1103/PhysRevD.43.4150.2} {\bibfield  {journal} {\bibinfo  {journal} {Phys.
  Rev.}\ }\textbf {\bibinfo {volume} {D42}},\ \bibinfo {pages} {354} (\bibinfo
  {year} {1990})},\ \bibinfo {note} {[Erratum: Phys.
  Rev.D43,4150(1991)]}\BibitemShut {NoStop}%
\bibitem [{Note1()}]{Note1}%
  \BibitemOpen
  \bibinfo {note} {Cosmic superstrings~\cite {Polchinski:2004yav}, the
  analogues of cosmic strings arising from string theory, are characterized
  also by the intercommutation probability which can take values between
  $10^{-3}$ and 1 for fundamental superstrings (F-strings) and between
  $10^{-1}$ and 1 for D-branes extended in one macroscopic dimension
  (D-strings). In our present study we concentrate on field theoretical
  objects~\cite {Sakellariadou:2009ev}, and in particular Nambu-Goto strings
  with intercommutation probability of order one.}\BibitemShut {Stop}%
\bibitem [{\citenamefont {Bennett}\ and\ \citenamefont
  {Bouchet}(1988)}]{BB1988}%
  \BibitemOpen
  \bibfield  {author} {\bibinfo {author} {\bibfnamefont {D.~P.}\ \bibnamefont
  {Bennett}}\ and\ \bibinfo {author} {\bibfnamefont {F.~R.}\ \bibnamefont
  {Bouchet}},\ }\href {\doibase 10.1103/PhysRevLett.60.257} {\bibfield
  {journal} {\bibinfo  {journal} {Phys.Rev.Lett.}\ }\textbf {\bibinfo {volume}
  {60}},\ \bibinfo {pages} {257} (\bibinfo {year} {1988})}\BibitemShut
  {NoStop}%
\bibitem [{\citenamefont {Allen}\ and\ \citenamefont
  {Shellard}(1990)}]{Allen:1990tv}%
  \BibitemOpen
  \bibfield  {author} {\bibinfo {author} {\bibfnamefont {B.}~\bibnamefont
  {Allen}}\ and\ \bibinfo {author} {\bibfnamefont {E.}~\bibnamefont
  {Shellard}},\ }\href {\doibase 10.1103/PhysRevLett.64.119} {\bibfield
  {journal} {\bibinfo  {journal} {Phys. Rev. Lett.}\ }\textbf {\bibinfo
  {volume} {64}},\ \bibinfo {pages} {119} (\bibinfo {year} {1990})}\BibitemShut
  {NoStop}%
\bibitem [{\citenamefont {Sakellariadou}\ and\ \citenamefont
  {Vilenkin}(1990)}]{Sakellariadou:1990nd}%
  \BibitemOpen
  \bibfield  {author} {\bibinfo {author} {\bibfnamefont {M.}~\bibnamefont
  {Sakellariadou}}\ and\ \bibinfo {author} {\bibfnamefont {A.}~\bibnamefont
  {Vilenkin}},\ }\href {\doibase 10.1103/PhysRevD.42.349} {\bibfield  {journal}
  {\bibinfo  {journal} {Phys. Rev. D}\ }\textbf {\bibinfo {volume} {42}},\
  \bibinfo {pages} {349} (\bibinfo {year} {1990})}\BibitemShut {NoStop}%
\bibitem [{\citenamefont {Vilenkin}(1981)}]{Vilenkin:1981bx}%
  \BibitemOpen
  \bibfield  {author} {\bibinfo {author} {\bibfnamefont {A.}~\bibnamefont
  {Vilenkin}},\ }\href {\doibase 10.1016/0370-2693(81)91144-8} {\bibfield
  {journal} {\bibinfo  {journal} {Phys.Lett.}\ }\textbf {\bibinfo {volume} {B
  107}},\ \bibinfo {pages} {47} (\bibinfo {year} {1981})}\BibitemShut {NoStop}%
\bibitem [{\citenamefont {Hogan}\ and\ \citenamefont
  {Rees}(1984)}]{Hogan:1984is}%
  \BibitemOpen
  \bibfield  {author} {\bibinfo {author} {\bibfnamefont {C.}~\bibnamefont
  {Hogan}}\ and\ \bibinfo {author} {\bibfnamefont {M.}~\bibnamefont {Rees}},\
  }\href {\doibase 10.1038/311109a0} {\bibfield  {journal} {\bibinfo  {journal}
  {Nature}\ }\textbf {\bibinfo {volume} {311}},\ \bibinfo {pages} {109}
  (\bibinfo {year} {1984})}\BibitemShut {NoStop}%
\bibitem [{Note2()}]{Note2}%
  \BibitemOpen
  \bibinfo {note} {Super-horizon cosmic strings also emit gravitational waves,
  due to their small-scale structure resulting from string
  intercommutations~\cite
  {Sakellariadou:1990ne,Hindmarsh:1990xi,Matsui:2016xnp}.}\BibitemShut {Stop}%
\bibitem [{\citenamefont {Damour}\ and\ \citenamefont
  {Vilenkin}(2000)}]{Damour:2000wa}%
  \BibitemOpen
  \bibfield  {author} {\bibinfo {author} {\bibfnamefont {T.}~\bibnamefont
  {Damour}}\ and\ \bibinfo {author} {\bibfnamefont {A.}~\bibnamefont
  {Vilenkin}},\ }\href {\doibase 10.1103/PhysRevLett.85.3761} {\bibfield
  {journal} {\bibinfo  {journal} {Phys.Rev.Lett.}\ }\textbf {\bibinfo {volume}
  {85}},\ \bibinfo {pages} {3761} (\bibinfo {year} {2000})},\ \Eprint
  {http://arxiv.org/abs/gr-qc/0004075} {arXiv:gr-qc/0004075 [gr-qc]}
  \BibitemShut {NoStop}%
\bibitem [{\citenamefont {Damour}\ and\ \citenamefont
  {Vilenkin}(2001)}]{Damour:2001bk}%
  \BibitemOpen
  \bibfield  {author} {\bibinfo {author} {\bibfnamefont {T.}~\bibnamefont
  {Damour}}\ and\ \bibinfo {author} {\bibfnamefont {A.}~\bibnamefont
  {Vilenkin}},\ }\href {\doibase 10.1103/PhysRevD.64.064008} {\bibfield
  {journal} {\bibinfo  {journal} {Phys.Rev.}\ }\textbf {\bibinfo {volume} {D
  64}},\ \bibinfo {pages} {064008} (\bibinfo {year} {2001})},\ \Eprint
  {http://arxiv.org/abs/gr-qc/0104026} {arXiv:gr-qc/0104026 [gr-qc]}
  \BibitemShut {NoStop}%
\bibitem [{\citenamefont {Damour}\ and\ \citenamefont
  {Vilenkin}(2005)}]{Damour:2004kw}%
  \BibitemOpen
  \bibfield  {author} {\bibinfo {author} {\bibfnamefont {T.}~\bibnamefont
  {Damour}}\ and\ \bibinfo {author} {\bibfnamefont {A.}~\bibnamefont
  {Vilenkin}},\ }\href {\doibase 10.1103/PhysRevD.71.063510} {\bibfield
  {journal} {\bibinfo  {journal} {Phys.Rev.}\ }\textbf {\bibinfo {volume} {D
  71}},\ \bibinfo {pages} {063510} (\bibinfo {year} {2005})},\ \Eprint
  {http://arxiv.org/abs/hep-th/0410222} {arXiv:hep-th/0410222 [hep-th]}
  \BibitemShut {NoStop}%
\bibitem [{Note3()}]{Note3}%
  \BibitemOpen
  \bibinfo {note} {We also include the so-called pseudocusps~\cite
  {Stott:2016loe}, defined as cuspy features moving with a velocity close to
  the speed of light.}\BibitemShut {Stop}%
\bibitem [{\citenamefont {Siemens}\ \emph {et~al.}(2007)\citenamefont
  {Siemens}, \citenamefont {Mandic},\ and\ \citenamefont
  {Creighton}}]{Siemens:2006yp}%
  \BibitemOpen
  \bibfield  {author} {\bibinfo {author} {\bibfnamefont {X.}~\bibnamefont
  {Siemens}}, \bibinfo {author} {\bibfnamefont {V.}~\bibnamefont {Mandic}}, \
  and\ \bibinfo {author} {\bibfnamefont {J.}~\bibnamefont {Creighton}},\ }\href
  {\doibase 10.1103/PhysRevLett.98.111101} {\bibfield  {journal} {\bibinfo
  {journal} {Phys. Rev. Lett.}\ }\textbf {\bibinfo {volume} {98}},\ \bibinfo
  {pages} {111101} (\bibinfo {year} {2007})},\ \Eprint
  {http://arxiv.org/abs/astro-ph/0610920} {arXiv:astro-ph/0610920 [astro-ph]}
  \BibitemShut {NoStop}%
\bibitem [{\citenamefont {Blanco-Pillado}\ \emph {et~al.}(2014)\citenamefont
  {Blanco-Pillado}, \citenamefont {Olum},\ and\ \citenamefont
  {Shlaer}}]{Blanco-Pillado:2013qja}%
  \BibitemOpen
  \bibfield  {author} {\bibinfo {author} {\bibfnamefont {J.~J.}\ \bibnamefont
  {Blanco-Pillado}}, \bibinfo {author} {\bibfnamefont {K.~D.}\ \bibnamefont
  {Olum}}, \ and\ \bibinfo {author} {\bibfnamefont {B.}~\bibnamefont
  {Shlaer}},\ }\href {\doibase 10.1103/PhysRevD.89.023512} {\bibfield
  {journal} {\bibinfo  {journal} {Phys. Rev.}\ }\textbf {\bibinfo {volume}
  {D89}},\ \bibinfo {pages} {023512} (\bibinfo {year} {2014})},\ \Eprint
  {http://arxiv.org/abs/1309.6637} {arXiv:1309.6637 [astro-ph.CO]} \BibitemShut
  {NoStop}%
\bibitem [{\citenamefont {Lorenz}\ \emph {et~al.}(2010)\citenamefont {Lorenz},
  \citenamefont {Ringeval},\ and\ \citenamefont
  {Sakellariadou}}]{Lorenz:2010sm}%
  \BibitemOpen
  \bibfield  {author} {\bibinfo {author} {\bibfnamefont {L.}~\bibnamefont
  {Lorenz}}, \bibinfo {author} {\bibfnamefont {C.}~\bibnamefont {Ringeval}}, \
  and\ \bibinfo {author} {\bibfnamefont {M.}~\bibnamefont {Sakellariadou}},\
  }\href {\doibase 10.1088/1475-7516/2010/10/003} {\bibfield  {journal}
  {\bibinfo  {journal} {JCAP}\ }\textbf {\bibinfo {volume} {1010}},\ \bibinfo
  {pages} {003} (\bibinfo {year} {2010})},\ \Eprint
  {http://arxiv.org/abs/1006.0931} {arXiv:1006.0931 [astro-ph.CO]} \BibitemShut
  {NoStop}%
\bibitem [{\citenamefont {Abbott}\ \emph {et~al.}()\citenamefont {Abbott} \emph
  {et~al.}}]{O3_isotropic}%
  \BibitemOpen
  \bibfield  {author} {\bibinfo {author} {\bibfnamefont {B.}~\bibnamefont
  {Abbott}} \emph {et~al.} (\bibinfo {collaboration} {LIGO Scientific
  Collaboration, Virgo}),\ }\href {https://dcc.ligo.org/LIGO-P2000314/public}
  {\bibinfo  {journal} {LIGO-DCC: P2000314}\ }\BibitemShut {NoStop}%
\bibitem [{\citenamefont {Abbott}\ \emph {et~al.}(2018)\citenamefont {Abbott}
  \emph {et~al.}}]{Abbott:2017mem}%
  \BibitemOpen
\bibfield  {journal} {  }\bibfield  {author} {\bibinfo {author} {\bibfnamefont
  {B.}~\bibnamefont {Abbott}} \emph {et~al.} (\bibinfo {collaboration} {LIGO
  Scientific Collaboration, Virgo}),\ }\href {\doibase
  10.1103/PhysRevD.97.102002} {\bibfield  {journal} {\bibinfo  {journal} {Phys.
  Rev. D}\ }\textbf {\bibinfo {volume} {97}},\ \bibinfo {pages} {102002}
  (\bibinfo {year} {2018})},\ \Eprint {http://arxiv.org/abs/1712.01168}
  {arXiv:1712.01168 [gr-qc]} \BibitemShut {NoStop}%
\bibitem [{\citenamefont {{Ade}}\ \emph {et~al.}(2016)\citenamefont {{Ade}}
  \emph {et~al.}}]{Planck_2015}%
  \BibitemOpen
  \bibfield  {author} {\bibinfo {author} {\bibfnamefont {P.~A.~R.}\
  \bibnamefont {{Ade}}} \emph {et~al.},\ }\href {\doibase
  10.1051/0004-6361/201525830} {\bibfield  {journal} {\bibinfo  {journal}
  {Astron. Astrophys.}\ }\textbf {\bibinfo {volume} {594}},\ \bibinfo {eid}
  {A13} (\bibinfo {year} {2016})}\BibitemShut {NoStop}%
\bibitem [{\citenamefont {Jenkins}\ and\ \citenamefont
  {Sakellariadou}(2018)}]{Jenkins:2018lvb}%
  \BibitemOpen
  \bibfield  {author} {\bibinfo {author} {\bibfnamefont {A.~C.}\ \bibnamefont
  {Jenkins}}\ and\ \bibinfo {author} {\bibfnamefont {M.}~\bibnamefont
  {Sakellariadou}},\ }\href {\doibase 10.1103/PhysRevD.98.063509} {\bibfield
  {journal} {\bibinfo  {journal} {Phys. Rev.}\ }\textbf {\bibinfo {volume}
  {D98}},\ \bibinfo {pages} {063509} (\bibinfo {year} {2018})},\ \Eprint
  {http://arxiv.org/abs/1802.06046} {arXiv:1802.06046 [astro-ph.CO]}
  \BibitemShut {NoStop}%
\bibitem [{\citenamefont {Gradshteyn}\ and\ \citenamefont
  {Ryzhik}(2007)}]{gradshteyn2007}%
  \BibitemOpen
  \bibfield  {author} {\bibinfo {author} {\bibfnamefont {I.~S.}\ \bibnamefont
  {Gradshteyn}}\ and\ \bibinfo {author} {\bibfnamefont {I.~M.}\ \bibnamefont
  {Ryzhik}},\ }\href@noop {} {\emph {\bibinfo {title} {Table of integrals,
  series, and products}}},\ \bibinfo {edition} {seventh}\ ed.\ (\bibinfo
  {publisher} {Elsevier/Academic Press, Amsterdam},\ \bibinfo {year} {2007})\
  pp.\ \bibinfo {pages} {xlviii+1171},\ \bibinfo {note} {translated from the
  Russian, Translation edited and with a preface by Alan Jeffrey and Daniel
  Zwillinger, With one CD-ROM (Windows, Macintosh and UNIX)}\BibitemShut
  {NoStop}%
\bibitem [{\citenamefont {Binetruy}\ \emph {et~al.}(2009)\citenamefont
  {Binetruy}, \citenamefont {Bohe}, \citenamefont {Hertog},\ and\ \citenamefont
  {Steer}}]{Binetruy:2009vt}%
  \BibitemOpen
  \bibfield  {author} {\bibinfo {author} {\bibfnamefont {P.}~\bibnamefont
  {Binetruy}}, \bibinfo {author} {\bibfnamefont {A.}~\bibnamefont {Bohe}},
  \bibinfo {author} {\bibfnamefont {T.}~\bibnamefont {Hertog}}, \ and\ \bibinfo
  {author} {\bibfnamefont {D.~A.}\ \bibnamefont {Steer}},\ }\href {\doibase
  10.1103/PhysRevD.80.123510} {\bibfield  {journal} {\bibinfo  {journal} {Phys.
  Rev. D}\ }\textbf {\bibinfo {volume} {80}},\ \bibinfo {pages} {123510}
  (\bibinfo {year} {2009})},\ \Eprint {http://arxiv.org/abs/0907.4522}
  {arXiv:0907.4522 [hep-th]} \BibitemShut {NoStop}%
\bibitem [{\citenamefont {Ringeval}\ \emph {et~al.}(2007)\citenamefont
  {Ringeval}, \citenamefont {Sakellariadou},\ and\ \citenamefont
  {Bouchet}}]{Ringeval:2005kr}%
  \BibitemOpen
  \bibfield  {author} {\bibinfo {author} {\bibfnamefont {C.}~\bibnamefont
  {Ringeval}}, \bibinfo {author} {\bibfnamefont {M.}~\bibnamefont
  {Sakellariadou}}, \ and\ \bibinfo {author} {\bibfnamefont {F.}~\bibnamefont
  {Bouchet}},\ }\href {\doibase 10.1088/1475-7516/2007/02/023} {\bibfield
  {journal} {\bibinfo  {journal} {JCAP}\ }\textbf {\bibinfo {volume} {0702}},\
  \bibinfo {pages} {023} (\bibinfo {year} {2007})},\ \Eprint
  {http://arxiv.org/abs/astro-ph/0511646} {arXiv:astro-ph/0511646 [astro-ph]}
  \BibitemShut {NoStop}%
\bibitem [{\citenamefont {Auclair}\ \emph {et~al.}(2019)\citenamefont
  {Auclair}, \citenamefont {Ringeval}, \citenamefont {Sakellariadou},\ and\
  \citenamefont {Steer}}]{Auclair:2019zoz}%
  \BibitemOpen
  \bibfield  {author} {\bibinfo {author} {\bibfnamefont {P.}~\bibnamefont
  {Auclair}}, \bibinfo {author} {\bibfnamefont {C.}~\bibnamefont {Ringeval}},
  \bibinfo {author} {\bibfnamefont {M.}~\bibnamefont {Sakellariadou}}, \ and\
  \bibinfo {author} {\bibfnamefont {D.}~\bibnamefont {Steer}},\ }\href
  {\doibase 10.1088/1475-7516/2019/06/015} {\bibfield  {journal} {\bibinfo
  {journal} {JCAP}\ }\textbf {\bibinfo {volume} {06}},\ \bibinfo {pages} {015}
  (\bibinfo {year} {2019})},\ \Eprint {http://arxiv.org/abs/1903.06685}
  {arXiv:1903.06685 [astro-ph.CO]} \BibitemShut {NoStop}%
\bibitem [{Note4()}]{Note4}%
  \BibitemOpen
  \bibinfo {note} {See Supplemental Material for more descriptions on the
  cosmic string loop distributions, and the burst analysis
  pipeline}\BibitemShut {NoStop}%
\bibitem [{\citenamefont {Allen}\ and\ \citenamefont
  {Shellard}(1992)}]{Allen:1991bk}%
  \BibitemOpen
  \bibfield  {author} {\bibinfo {author} {\bibfnamefont {B.}~\bibnamefont
  {Allen}}\ and\ \bibinfo {author} {\bibfnamefont {E.}~\bibnamefont
  {Shellard}},\ }\href {\doibase 10.1103/PhysRevD.45.1898} {\bibfield
  {journal} {\bibinfo  {journal} {Phys. Rev. D}\ }\textbf {\bibinfo {volume}
  {45}},\ \bibinfo {pages} {1898} (\bibinfo {year} {1992})}\BibitemShut
  {NoStop}%
\bibitem [{\citenamefont {Christensen}(2019)}]{Christensen:2018iqi}%
  \BibitemOpen
  \bibfield  {author} {\bibinfo {author} {\bibfnamefont {N.}~\bibnamefont
  {Christensen}},\ }\href {\doibase 10.1088/1361-6633/aae6b5} {\bibfield
  {journal} {\bibinfo  {journal} {Rept. Prog. Phys.}\ }\textbf {\bibinfo
  {volume} {82}},\ \bibinfo {pages} {016903} (\bibinfo {year} {2019})},\
  \Eprint {http://arxiv.org/abs/1811.08797} {arXiv:1811.08797 [gr-qc]}
  \BibitemShut {NoStop}%
\bibitem [{\citenamefont {Olmez}\ \emph {et~al.}(2010)\citenamefont {Olmez},
  \citenamefont {Mandic},\ and\ \citenamefont {Siemens}}]{Olmez:2010bi}%
  \BibitemOpen
  \bibfield  {author} {\bibinfo {author} {\bibfnamefont {S.}~\bibnamefont
  {Olmez}}, \bibinfo {author} {\bibfnamefont {V.}~\bibnamefont {Mandic}}, \
  and\ \bibinfo {author} {\bibfnamefont {X.}~\bibnamefont {Siemens}},\ }\href
  {\doibase 10.1103/PhysRevD.81.104028} {\bibfield  {journal} {\bibinfo
  {journal} {Phys.Rev.}\ }\textbf {\bibinfo {volume} {D 81}},\ \bibinfo {pages}
  {104028} (\bibinfo {year} {2010})},\ \Eprint {http://arxiv.org/abs/1004.0890}
  {arXiv:1004.0890 [astro-ph.CO]} \BibitemShut {NoStop}%
\bibitem [{\citenamefont {Jenkins}\ and\ \citenamefont
  {Sakellariadou}(2020)}]{Jenkins:2020ctp}%
  \BibitemOpen
  \bibfield  {author} {\bibinfo {author} {\bibfnamefont {A.~C.}\ \bibnamefont
  {Jenkins}}\ and\ \bibinfo {author} {\bibfnamefont {M.}~\bibnamefont
  {Sakellariadou}},\ }\href@noop {} {\  (\bibinfo {year} {2020})},\ \Eprint
  {http://arxiv.org/abs/2006.16249} {arXiv:2006.16249 [astro-ph.CO]}
  \BibitemShut {NoStop}%
\bibitem [{\citenamefont {Thrane}\ and\ \citenamefont
  {Romano}(2013)}]{PhysRevD.88.124032}%
  \BibitemOpen
  \bibfield  {author} {\bibinfo {author} {\bibfnamefont {E.}~\bibnamefont
  {Thrane}}\ and\ \bibinfo {author} {\bibfnamefont {J.~D.}\ \bibnamefont
  {Romano}},\ }\href {\doibase 10.1103/PhysRevD.88.124032} {\bibfield
  {journal} {\bibinfo  {journal} {Phys. Rev. D}\ }\textbf {\bibinfo {volume}
  {88}},\ \bibinfo {pages} {124032} (\bibinfo {year} {2013})}\BibitemShut
  {NoStop}%
\bibitem [{\citenamefont {Abbott}\ \emph
  {et~al.}(2020{\natexlab{c}})\citenamefont {Abbott} \emph
  {et~al.}}]{Aasi:2013wya}%
  \BibitemOpen
  \bibfield  {author} {\bibinfo {author} {\bibfnamefont {B.}~\bibnamefont
  {Abbott}} \emph {et~al.} (\bibinfo {collaboration} {KAGRA, LIGO Scientific
  Collaboration, Virgo}),\ }\href {\doibase 10.1007/s41114-020-00026-9}
  {\bibfield  {journal} {\bibinfo  {journal} {Living Rev. Rel.}\ }\textbf
  {\bibinfo {volume} {23}},\ \bibinfo {pages} {3} (\bibinfo {year}
  {2020}{\natexlab{c}})},\ \Eprint {http://arxiv.org/abs/1304.0670}
  {arXiv:1304.0670 [gr-qc]} \BibitemShut {NoStop}%
\bibitem [{\citenamefont {Abbott}\ \emph
  {et~al.}(2019{\natexlab{a}})\citenamefont {Abbott} \emph
  {et~al.}}]{Abbott:2019prv}%
  \BibitemOpen
  \bibfield  {author} {\bibinfo {author} {\bibfnamefont {B.}~\bibnamefont
  {Abbott}} \emph {et~al.} (\bibinfo {collaboration} {LIGO Scientific
  Collaboration, Virgo}),\ }\href {\doibase 10.1103/PhysRevD.100.024017}
  {\bibfield  {journal} {\bibinfo  {journal} {Phys. Rev. D}\ }\textbf {\bibinfo
  {volume} {100}},\ \bibinfo {pages} {024017} (\bibinfo {year}
  {2019}{\natexlab{a}})},\ \Eprint {http://arxiv.org/abs/1905.03457}
  {arXiv:1905.03457 [gr-qc]} \BibitemShut {NoStop}%
\bibitem [{\citenamefont {Aasi}\ \emph {et~al.}(2014)\citenamefont {Aasi} \emph
  {et~al.}}]{Aasi:2013vna}%
  \BibitemOpen
  \bibfield  {author} {\bibinfo {author} {\bibfnamefont {J.}~\bibnamefont
  {Aasi}} \emph {et~al.} (\bibinfo {collaboration} {Virgo, LIGO Scientific
  Collaboration}),\ }\href {\doibase 10.1103/PhysRevLett.112.131101} {\bibfield
   {journal} {\bibinfo  {journal} {Phys. Rev. Lett.}\ }\textbf {\bibinfo
  {volume} {112}},\ \bibinfo {pages} {131101} (\bibinfo {year} {2014})},\
  \Eprint {http://arxiv.org/abs/1310.2384} {arXiv:1310.2384 [gr-qc]}
  \BibitemShut {NoStop}%
\bibitem [{\citenamefont {Cannon}(2008)}]{Cannon:2008zz}%
  \BibitemOpen
  \bibfield  {author} {\bibinfo {author} {\bibfnamefont {K.~C.}\ \bibnamefont
  {Cannon}},\ }\href {\doibase 10.1088/0264-9381/25/10/105024} {\bibfield
  {journal} {\bibinfo  {journal} {Class.Quant.Grav.}\ }\textbf {\bibinfo
  {volume} {25}},\ \bibinfo {pages} {105024} (\bibinfo {year}
  {2008})}\BibitemShut {NoStop}%
\bibitem [{\citenamefont {Abbott}\ \emph {et~al.}(2016)\citenamefont {Abbott}
  \emph {et~al.}}]{TheLIGOScientific:2016zmo}%
  \BibitemOpen
  \bibfield  {author} {\bibinfo {author} {\bibfnamefont {B.~P.}\ \bibnamefont
  {Abbott}} \emph {et~al.} (\bibinfo {collaboration} {Virgo, LIGO Scientific
  Collaboration}),\ }\href {\doibase 10.1088/0264-9381/33/13/134001} {\bibfield
   {journal} {\bibinfo  {journal} {Class. Quant. Grav.}\ }\textbf {\bibinfo
  {volume} {33}},\ \bibinfo {pages} {134001} (\bibinfo {year} {2016})},\
  \Eprint {http://arxiv.org/abs/1602.03844} {arXiv:1602.03844 [gr-qc]}
  \BibitemShut {NoStop}%
\bibitem [{\citenamefont {Cabero}\ \emph {et~al.}(2019)\citenamefont {Cabero}
  \emph {et~al.}}]{Cabero:2019orq}%
  \BibitemOpen
  \bibfield  {author} {\bibinfo {author} {\bibfnamefont {M.}~\bibnamefont
  {Cabero}} \emph {et~al.},\ }\href {\doibase 10.1088/1361-6382/ab2e14}
  {\bibfield  {journal} {\bibinfo  {journal} {Class. Quant. Grav.}\ }\textbf
  {\bibinfo {volume} {36}},\ \bibinfo {pages} {15} (\bibinfo {year} {2019})},\
  \Eprint {http://arxiv.org/abs/1901.05093} {arXiv:1901.05093
  [physics.ins-det]} \BibitemShut {NoStop}%
\bibitem [{\citenamefont {Brady}\ \emph {et~al.}(2004)\citenamefont {Brady},
  \citenamefont {Creighton},\ and\ \citenamefont {Wiseman}}]{Brady:2004gt}%
  \BibitemOpen
  \bibfield  {author} {\bibinfo {author} {\bibfnamefont {P.~R.}\ \bibnamefont
  {Brady}}, \bibinfo {author} {\bibfnamefont {J.~D.}\ \bibnamefont
  {Creighton}}, \ and\ \bibinfo {author} {\bibfnamefont {A.~G.}\ \bibnamefont
  {Wiseman}},\ }\href {\doibase 10.1088/0264-9381/21/20/020} {\bibfield
  {journal} {\bibinfo  {journal} {Class.Quant.Grav.}\ }\textbf {\bibinfo
  {volume} {21}},\ \bibinfo {pages} {S1775} (\bibinfo {year} {2004})},\ \Eprint
  {http://arxiv.org/abs/gr-qc/0405044} {arXiv:gr-qc/0405044 [gr-qc]}
  \BibitemShut {NoStop}%
\bibitem [{\citenamefont {Abbott}\ \emph {et~al.}(2017)\citenamefont {Abbott}
  \emph {et~al.}}]{TheLIGOScientific:2016dpb}%
  \BibitemOpen
  \bibfield  {author} {\bibinfo {author} {\bibfnamefont {B.~P.}\ \bibnamefont
  {Abbott}} \emph {et~al.} (\bibinfo {collaboration} {Virgo, LIGO Scientific
  Collaboration}),\ }\href {\doibase 10.1103/PhysRevLett.118.121101} {\bibfield
   {journal} {\bibinfo  {journal} {Phys. Rev. Lett.}\ }\textbf {\bibinfo
  {volume} {118}},\ \bibinfo {pages} {121101} (\bibinfo {year} {2017})},\
  \Eprint {http://arxiv.org/abs/1612.02029} {arXiv:1612.02029 [gr-qc]}
  \BibitemShut {NoStop}%
\bibitem [{\citenamefont {Abbott}\ \emph
  {et~al.}(2019{\natexlab{b}})\citenamefont {Abbott} \emph
  {et~al.}}]{LIGOScientific:2019vic}%
  \BibitemOpen
  \bibfield  {author} {\bibinfo {author} {\bibfnamefont {B.}~\bibnamefont
  {Abbott}} \emph {et~al.} (\bibinfo {collaboration} {LIGO Scientific
  Collaboration, Virgo}),\ }\href {\doibase 10.1103/PhysRevD.100.061101}
  {\bibfield  {journal} {\bibinfo  {journal} {Phys. Rev. D}\ }\textbf {\bibinfo
  {volume} {100}},\ \bibinfo {pages} {061101} (\bibinfo {year}
  {2019}{\natexlab{b}})},\ \Eprint {http://arxiv.org/abs/1903.02886}
  {arXiv:1903.02886 [gr-qc]} \BibitemShut {NoStop}%
\bibitem [{\citenamefont {Mandic}\ \emph {et~al.}(2012)\citenamefont {Mandic},
  \citenamefont {Thrane}, \citenamefont {Giampanis},\ and\ \citenamefont
  {Regimbau}}]{Mandic_2012}%
  \BibitemOpen
  \bibfield  {author} {\bibinfo {author} {\bibfnamefont {V.}~\bibnamefont
  {Mandic}}, \bibinfo {author} {\bibfnamefont {E.}~\bibnamefont {Thrane}},
  \bibinfo {author} {\bibfnamefont {S.}~\bibnamefont {Giampanis}}, \ and\
  \bibinfo {author} {\bibfnamefont {T.}~\bibnamefont {Regimbau}},\ }\href
  {\doibase 10.1103/physrevlett.109.171102} {\bibfield  {journal} {\bibinfo
  {journal} {Physical Review Letters}\ }\textbf {\bibinfo {volume} {109}}
  (\bibinfo {year} {2012}),\ 10.1103/physrevlett.109.171102}\BibitemShut
  {NoStop}%
\bibitem [{\citenamefont {Ade}\ \emph {et~al.}(2013)\citenamefont {Ade} \emph
  {et~al.}}]{Ade:2013xla}%
  \BibitemOpen
  \bibfield  {author} {\bibinfo {author} {\bibfnamefont {P.}~\bibnamefont
  {Ade}} \emph {et~al.} (\bibinfo {collaboration} {Planck Collaboration}),\
  }\href@noop {} {\  (\bibinfo {year} {2013})},\ \Eprint
  {http://arxiv.org/abs/1303.5085} {arXiv:1303.5085 [astro-ph.CO]} \BibitemShut
  {NoStop}%
\bibitem [{\citenamefont {Lizarraga}\ \emph {et~al.}(2016)\citenamefont
  {Lizarraga}, \citenamefont {Urrestilla}, \citenamefont {Daverio},
  \citenamefont {Hindmarsh},\ and\ \citenamefont {Kunz}}]{Lizarraga:2016onn}%
  \BibitemOpen
  \bibfield  {author} {\bibinfo {author} {\bibfnamefont {J.}~\bibnamefont
  {Lizarraga}}, \bibinfo {author} {\bibfnamefont {J.}~\bibnamefont
  {Urrestilla}}, \bibinfo {author} {\bibfnamefont {D.}~\bibnamefont {Daverio}},
  \bibinfo {author} {\bibfnamefont {M.}~\bibnamefont {Hindmarsh}}, \ and\
  \bibinfo {author} {\bibfnamefont {M.}~\bibnamefont {Kunz}},\ }\href {\doibase
  10.1088/1475-7516/2016/10/042} {\bibfield  {journal} {\bibinfo  {journal}
  {JCAP}\ }\textbf {\bibinfo {volume} {1610}},\ \bibinfo {pages} {042}
  (\bibinfo {year} {2016})},\ \Eprint {http://arxiv.org/abs/1609.03386}
  {arXiv:1609.03386 [astro-ph.CO]} \BibitemShut {NoStop}%
\bibitem [{\citenamefont {Henrot-Versille}\ \emph {et~al.}(2015)\citenamefont
  {Henrot-Versille} \emph {et~al.}}]{Henrot-Versille:2014jua}%
  \BibitemOpen
  \bibfield  {author} {\bibinfo {author} {\bibfnamefont {S.}~\bibnamefont
  {Henrot-Versille}} \emph {et~al.},\ }\href {\doibase
  10.1088/0264-9381/32/4/045003} {\bibfield  {journal} {\bibinfo  {journal}
  {Class. Quant. Grav.}\ }\textbf {\bibinfo {volume} {32}},\ \bibinfo {pages}
  {045003} (\bibinfo {year} {2015})},\ \Eprint {http://arxiv.org/abs/1408.5299}
  {arXiv:1408.5299 [astro-ph.CO]} \BibitemShut {NoStop}%
\bibitem [{\citenamefont {Lazanu}\ and\ \citenamefont
  {Shellard}(2015)}]{Lazanu:2014eya}%
  \BibitemOpen
  \bibfield  {author} {\bibinfo {author} {\bibfnamefont {A.}~\bibnamefont
  {Lazanu}}\ and\ \bibinfo {author} {\bibfnamefont {P.}~\bibnamefont
  {Shellard}},\ }\href {\doibase 10.1088/1475-7516/2015/02/024} {\bibfield
  {journal} {\bibinfo  {journal} {JCAP}\ }\textbf {\bibinfo {volume} {1502}},\
  \bibinfo {pages} {024} (\bibinfo {year} {2015})},\ \Eprint
  {http://arxiv.org/abs/1410.5046} {arXiv:1410.5046 [astro-ph.CO]} \BibitemShut
  {NoStop}%
\bibitem [{\citenamefont {Auclair}\ \emph {et~al.}(2020)\citenamefont {Auclair}
  \emph {et~al.}}]{Auclair:2019wcv}%
  \BibitemOpen
  \bibfield  {author} {\bibinfo {author} {\bibfnamefont {P.}~\bibnamefont
  {Auclair}} \emph {et~al.},\ }\href {\doibase 10.1088/1475-7516/2020/04/034}
  {\bibfield  {journal} {\bibinfo  {journal} {JCAP}\ }\textbf {\bibinfo
  {volume} {04}},\ \bibinfo {pages} {034} (\bibinfo {year} {2020})},\ \Eprint
  {http://arxiv.org/abs/1909.00819} {arXiv:1909.00819 [astro-ph.CO]}
  \BibitemShut {NoStop}%
\bibitem [{\citenamefont {Lasky}\ \emph {et~al.}(2016)\citenamefont {Lasky}
  \emph {et~al.}}]{Lasky:2015lej}%
  \BibitemOpen
  \bibfield  {author} {\bibinfo {author} {\bibfnamefont {P.~D.}\ \bibnamefont
  {Lasky}} \emph {et~al.},\ }\href {\doibase 10.1103/PhysRevX.6.011035}
  {\bibfield  {journal} {\bibinfo  {journal} {Phys. Rev.}\ }\textbf {\bibinfo
  {volume} {X6}},\ \bibinfo {pages} {011035} (\bibinfo {year} {2016})},\
  \Eprint {http://arxiv.org/abs/1511.05994} {arXiv:1511.05994 [astro-ph.CO]}
  \BibitemShut {NoStop}%
\bibitem [{\citenamefont {Auclair}(2020)}]{Auclair:2020oww}%
  \BibitemOpen
  \bibfield  {author} {\bibinfo {author} {\bibfnamefont {P.~G.}\ \bibnamefont
  {Auclair}},\ }\href {\doibase 10.1088/1475-7516/2020/11/050} {\bibfield
  {journal} {\bibinfo  {journal} {JCAP}\ }\textbf {\bibinfo {volume} {11}},\
  \bibinfo {pages} {050} (\bibinfo {year} {2020})},\ \Eprint
  {http://arxiv.org/abs/2009.00334} {arXiv:2009.00334 [astro-ph.CO]}
  \BibitemShut {NoStop}%
\bibitem [{\citenamefont {Ringeval}\ and\ \citenamefont
  {Suyama}(2017)}]{Ringeval:2017eww}%
  \BibitemOpen
  \bibfield  {author} {\bibinfo {author} {\bibfnamefont {C.}~\bibnamefont
  {Ringeval}}\ and\ \bibinfo {author} {\bibfnamefont {T.}~\bibnamefont
  {Suyama}},\ }\href {\doibase 10.1088/1475-7516/2017/12/027} {\bibfield
  {journal} {\bibinfo  {journal} {JCAP}\ }\textbf {\bibinfo {volume} {12}},\
  \bibinfo {pages} {027} (\bibinfo {year} {2017})},\ \Eprint
  {http://arxiv.org/abs/1709.03845} {arXiv:1709.03845 [astro-ph.CO]}
  \BibitemShut {NoStop}%
\bibitem [{\citenamefont {Pazouli}\ \emph {et~al.}(2020)\citenamefont
  {Pazouli}, \citenamefont {Avgoustidis},\ and\ \citenamefont
  {Copeland}}]{Pazouli:2020qmj}%
  \BibitemOpen
  \bibfield  {author} {\bibinfo {author} {\bibfnamefont {D.}~\bibnamefont
  {Pazouli}}, \bibinfo {author} {\bibfnamefont {A.}~\bibnamefont
  {Avgoustidis}}, \ and\ \bibinfo {author} {\bibfnamefont {E.~J.}\ \bibnamefont
  {Copeland}},\ }\href@noop {} {\  (\bibinfo {year} {2020})},\ \Eprint
  {http://arxiv.org/abs/2008.13693} {arXiv:2008.13693 [hep-th]} \BibitemShut
  {NoStop}%
\bibitem [{\citenamefont {Polchinski}(2004)}]{Polchinski:2004yav}%
  \BibitemOpen
  \bibfield  {author} {\bibinfo {author} {\bibfnamefont {J.}~\bibnamefont
  {Polchinski}},\ }\href {\doibase 10.1142/S0217751X05026686} {\bibfield
  {journal} {\bibinfo  {journal} {AIP Conf. Proc.}\ }\textbf {\bibinfo {volume}
  {743}},\ \bibinfo {pages} {331} (\bibinfo {year} {2004})},\ \Eprint
  {http://arxiv.org/abs/hep-th/0410082} {arXiv:hep-th/0410082} \BibitemShut
  {NoStop}%
\bibitem [{\citenamefont {Sakellariadou}(2009)}]{Sakellariadou:2009ev}%
  \BibitemOpen
  \bibfield  {author} {\bibinfo {author} {\bibfnamefont {M.}~\bibnamefont
  {Sakellariadou}},\ }\href {\doibase 10.1016/j.nuclphysbps.2009.07.046}
  {\bibfield  {journal} {\bibinfo  {journal} {Nucl. Phys. B Proc. Suppl.}\
  }\textbf {\bibinfo {volume} {192-193}},\ \bibinfo {pages} {68} (\bibinfo
  {year} {2009})},\ \Eprint {http://arxiv.org/abs/0902.0569} {arXiv:0902.0569
  [hep-th]} \BibitemShut {NoStop}%
\bibitem [{\citenamefont {Hindmarsh}(1990)}]{Hindmarsh:1990xi}%
  \BibitemOpen
  \bibfield  {author} {\bibinfo {author} {\bibfnamefont {M.}~\bibnamefont
  {Hindmarsh}},\ }\href {\doibase 10.1016/0370-2693(90)90226-V} {\bibfield
  {journal} {\bibinfo  {journal} {Phys. Lett.}\ }\textbf {\bibinfo {volume}
  {B251}},\ \bibinfo {pages} {28} (\bibinfo {year} {1990})}\BibitemShut
  {NoStop}%
\bibitem [{\citenamefont {Matsui}\ \emph {et~al.}(2016)\citenamefont {Matsui},
  \citenamefont {Horiguchi}, \citenamefont {Nitta},\ and\ \citenamefont
  {Kuroyanagi}}]{Matsui:2016xnp}%
  \BibitemOpen
  \bibfield  {author} {\bibinfo {author} {\bibfnamefont {Y.}~\bibnamefont
  {Matsui}}, \bibinfo {author} {\bibfnamefont {K.}~\bibnamefont {Horiguchi}},
  \bibinfo {author} {\bibfnamefont {D.}~\bibnamefont {Nitta}}, \ and\ \bibinfo
  {author} {\bibfnamefont {S.}~\bibnamefont {Kuroyanagi}},\ }\href {\doibase
  10.1088/1475-7516/2016/11/005} {\bibfield  {journal} {\bibinfo  {journal}
  {JCAP}\ }\textbf {\bibinfo {volume} {1611}},\ \bibinfo {pages} {005}
  (\bibinfo {year} {2016})},\ \Eprint {http://arxiv.org/abs/1605.08768}
  {arXiv:1605.08768 [astro-ph.CO]} \BibitemShut {NoStop}%
\bibitem [{\citenamefont {Stott}\ \emph {et~al.}(2017)\citenamefont {Stott},
  \citenamefont {Elghozi},\ and\ \citenamefont
  {Sakellariadou}}]{Stott:2016loe}%
  \BibitemOpen
  \bibfield  {author} {\bibinfo {author} {\bibfnamefont {M.~J.}\ \bibnamefont
  {Stott}}, \bibinfo {author} {\bibfnamefont {T.}~\bibnamefont {Elghozi}}, \
  and\ \bibinfo {author} {\bibfnamefont {M.}~\bibnamefont {Sakellariadou}},\
  }\href {\doibase 10.1103/PhysRevD.96.023533} {\bibfield  {journal} {\bibinfo
  {journal} {Phys. Rev. D}\ }\textbf {\bibinfo {volume} {96}},\ \bibinfo
  {pages} {023533} (\bibinfo {year} {2017})},\ \Eprint
  {http://arxiv.org/abs/1612.07599} {arXiv:1612.07599 [hep-th]} \BibitemShut
  {NoStop}%
\end{thebibliography}%


%
